%% file: main.tex
\documentclass[lettersize,journal]{IEEEtran}
\usepackage{times,amsmath,epsfig}
\usepackage[subtle=yes,tracking=normal]{savetrees} 
\usepackage{graphicx}
\usepackage{pgfplots}
\pgfplotsset{compat=newest}
\usepackage{balance}  
\usepackage{verbatim}

\usepackage[shortlabels]{enumitem}
\usepackage{tikz} 
\usetikzlibrary{positioning}
\usetikzlibrary{patterns}
\usepackage{color,soul}
\usepackage{booktabs}
\usepackage{url}
\usepackage{subcaption}
\usepackage{pgfplotstable}
\usetikzlibrary{pgfplots.groupplots}
\usepackage{stfloats}
\usepackage{ragged2e}

\usepackage{hyperref}
\hypersetup{
  colorlinks=true,
  linkcolor=blue,
  urlcolor=cyan,
}

\input{commands.tex}

\input{tables_funcs.tex}

\renewcommand{\IEEEbibitemsep}{0pt plus 0.5pt}
\makeatletter
\IEEEtriggercmd{\reset@font\normalfont\fontsize{7.5pt}{8.40pt}\selectfont}
\makeatother
\IEEEtriggeratref{1}

\ifCLASSOPTIONcompsoc
  \usepackage[nocompress]{cite}
\else
  \usepackage{cite}
\fi

\newcommand\copyrighttext{%
  \footnotesize \textcopyright 2024 IEEE. Personal use of this material is permitted.
  Permission from IEEE must be obtained for all other uses, in any current or future 
  media, including reprinting/republishing this material for advertising or promotional 
  purposes, creating new collective works, for resale or redistribution to servers or 
  lists, or reuse of any copyrighted component of this work in other works. 
  DOI: \href{https://doi.org/10.1109/TKDE.2024.3381192}{TKDE.2024.3381192}}
\newcommand\copyrightnotice{%
\begin{tikzpicture}[remember picture,overlay]
\node[anchor=south,yshift=10pt] at (current page.south) {\fbox{\parbox{\dimexpr\textwidth-\fboxsep-\fboxrule\relax}{\copyrighttext}}};
\end{tikzpicture}%
}

\begin{document}

\title{Diba: A Re-configurable Stream Processor}

\author{
	{Mohammadreza Najafi{\small $^1$}, Thamir M. Qadah{\small $^{2}$}, Mohammad Sadoghi{\small $^{3}$}, Hans-Arno Jacobsen{\small $^{4}$}, IEEE Fellow}\\
	\fontsize{10}{10}\selectfont\rmfamily\itshape Technical University of Munich$^{1}$,
	\fontsize{10}{10}\selectfont\rmfamily\itshape Umm Al-Qura University$^{2}$ \\
	\fontsize{10}{10}\selectfont\rmfamily\itshape University of California - Davis$^{3}$,
	\fontsize{10}{10}\selectfont\rmfamily\itshape University of Toronto$^{4}$ 
 \\
 \vspace{5mm}
 \textbf{This paper is accepted and published by IEEE TKDE in September 2024. DOI: \href{https://doi.org/10.1109/TKDE.2024.3381192}{TKDE.2024.3381192}}
}
\vspace{-5mm}

\IEEEtitleabstractindextext{
	
\input{abstract.tex}


}

\maketitle

\copyrightnotice

\IEEEdisplaynontitleabstractindextext
\IEEEpeerreviewmaketitle

\input{introduction.tex}

\input{related_work.tex}

\input{overview.tex}
\input{experimentalresults.tex}

\input{conclusions.tex}

\ifCLASSOPTIONcompsoc
  \section*{Acknowledgments}
\else
  \section*{Acknowledgment}
\fi

This research was in part supported by the Alexander von Humboldt
Foundation.

\ifCLASSOPTIONcaptionsoff
  \newpage
\fi

\bibliographystyle{IEEEtran}
\bibliography{references}   

\begin{IEEEbiography}
	[{\includegraphics[width=1in,height=1.25in,clip,keepaspectratio]{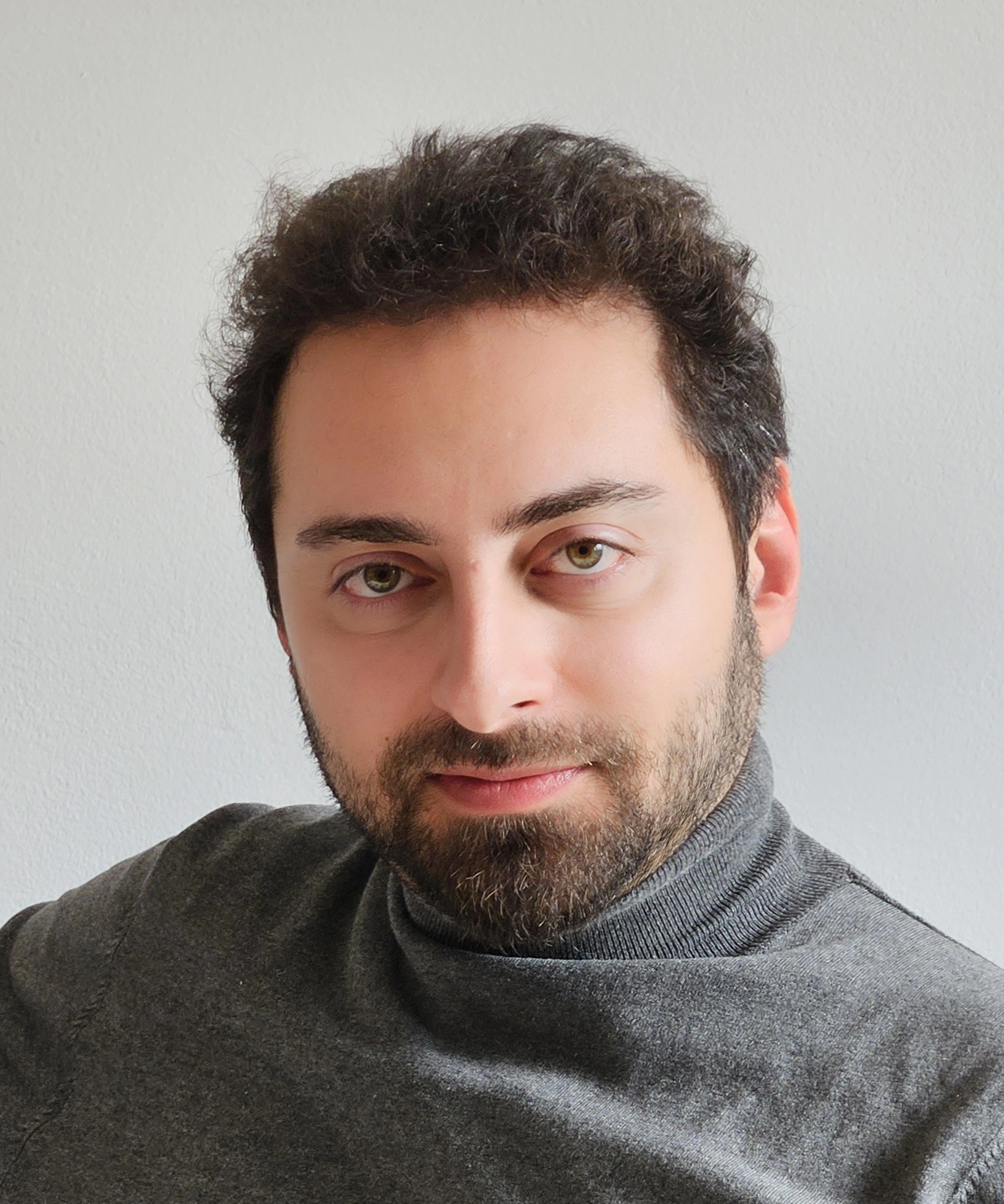}}]{Mohammadreza
          Najafi} a distinguished professional renowned for expertise
        in hardware design, system architecture, and pioneering
        research, embarked on his academic journey with Bachelor's and
        Master's degrees in Computer Engineering and Computer
        Architecture from the esteemed University of Tehran. Under the
        adept mentorship of Prof. Zainalabedin Navabi, he honed skills
        in advanced processor and cache architectures, network-on-chip
        design, and soft error evaluation techniques. At the Technical
        University of Munich, he attained a Doctorate with the highest
        honors, focusing on hardware-accelerated stream processing
        under the guidance of Prof. Hans-Arno Jacobsen and
        Prof. Mohammad Sadoghi, spearheading innovative advancements
        in real-time data processing. Currently, he holds a
        distinguished position at Synopsys, where he oversees a
        diverse portfolio of processor IPs, ranging from ultra-low
        power to high-performance processors, VLIW vector processors,
        neural processing units, and a variety of state-of-the-art
        tools, fostering a culture of innovation and excellence
        throughout the organization.
\end{IEEEbiography}

\begin{IEEEbiography}
	[{\includegraphics[width=1in,height=1.25in,trim={0 7mm 0
                0},clip,keepaspectratio]{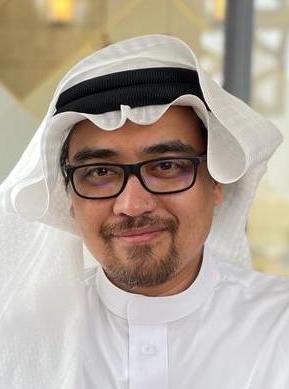}}]{Thamir M. Qadah}
        received his Ph.D. degree from Purdue University, West
        Lafayette, Indiana, in 2021. He works as an assistant
        professor at the College of Computing, Umm Al-Qura University
        in Makkah, Saudi Arabia. His research interests lie in
        designing and implementing secure, dependable, and
        high-performance software systems using modern hardware
        technologies, cloud infrastructures, and advanced AI and ML
        techniques. He has been awarded the Best Paper Award in
        Middleware '18 for his research on queue-oriented transaction
        processing, and he has since been expanding his research into
        building AI technologies and systems. Since 2015, he has been
        a reviewer for top-tier conferences such as SIGMOD, VLDB,
        ICDE, ICDCS, ATC, EDBT, Middleware, and CIKM. He also served
        as a journal reviewer for the IEEE (TKDE, TPDS, TSC, and
        Access), Springer DAPD, and as a committee member of the
        artifact evaluation committee for ASPLOS, OSDI, and SOSP. He
        has been an IEEE member since 2005.
\end{IEEEbiography}

\begin{IEEEbiography}
[{\includegraphics[width=1in,height=1.25in,trim={0 7mm 0
        0},clip,keepaspectratio]{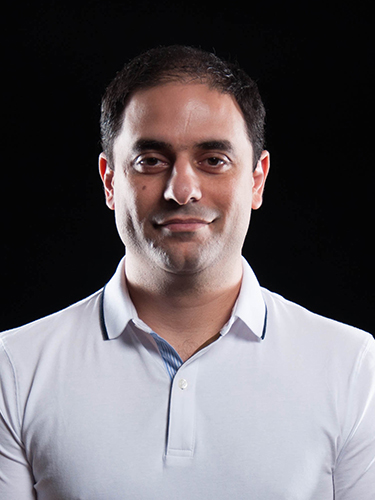}}]{Mohammad Sadoghi} is an
Associate Professor of Computer Science at the University of
California, Davis. Formerly, he was an Assistant Professor at Purdue
University and Research Staff Member at IBM T.J. Watson Research
Center.  He received his Ph.D. from the University of Toronto in
2013. His research is focused to pioneer a resilient data platform at
scale. He has over 100 publications and has filed 35 U.S. patents.  He
is serving in the Steering Committee of ACM/IFIP Middleware and
International Symposium on Foundations and Applications of Blockchain
(FAB), and served as the Lead-Guest Editor for Springer Distributed
and Parallel Databases: Special Issue on Blockchain. He has served as
the Program Co-Chair for Middleware 2025, Associate PC Chair for IEEE
ICDE 2023; Program Vice Co-Chair for IEEE ICDCS'21; Program Vice
Co-Chair for IEEE Big Data'21; General Chair of FAB'21; PC Co-Chair
for ACM DEBS'20; General Co-chair of ACM/IFIP Middleware'19; and PC
Chair (Industry Track) for ACM DEBS'17.  He has co-authored two books
as part of the Morgan \& Claypool Synthesis Lectures on Data Management (now
Springer): ``Transaction Processing on Modern Hardware'' and
``Fault-tolerant Distributed Transactions on Blockchain''.  He has
further co-authored a book published by Foundations and Trends® in
Databases, entitled ``Consensus in Data Management: From Distributed
Commit to Blockchain.''
\end{IEEEbiography}

\begin{IEEEbiography}
	[{\includegraphics[width=1in,height=1.25in,trim={0 7mm 0
                0},clip,keepaspectratio]{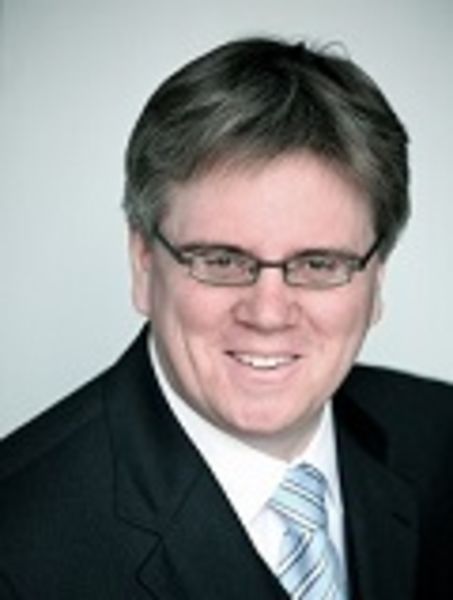}}]{Hans-Arno
          Jacobsen} is a pioneering researcher whose work lies at the
        interface among computer science, computer engineering and
        information systems. He holds numerous patents and has been
        involved in important industrial developments with partners
        like Bell Canada, Computer Associates, IBM, Yahoo! and Sun
        Microsystems. His principal areas of research include
        distributed systems, middleware systems, event processing and
        data management.  He has also explored the integration of
        hardware components, such as FPGAs, into middleware
        architectures and data management systems. After studying and
        completing his doctorate in Germany, France and the U.S.,
        Dr. Jacobsen engaged in post-doctoral research at INRIA near
        Paris before moving to the University of Toronto in 2001.
\end{IEEEbiography}

\end{document}

%% file: commands.tex
\definecolor{color7}{RGB}{0,0,0}
\definecolor{color6}{RGB}{200,190,20}
\definecolor{color2}{RGB}{230,159,0}
\definecolor{color3}{RGB}{86,180,233}
\definecolor{color4}{RGB}{0,158,115}
\definecolor{color5}{RGB}{0,114,178}
\definecolor{color1}{RGB}{213,94,0}

\pgfplotscreateplotcyclelist{colorbrewer-RYB}{
	{color1!50!black,fill=color1},
	{color2!50!black,fill=color2},
	{color3!50!black,fill=color3},
	{color4!50!black,fill=color4},
	{color5!50!black,fill=color5},
	{color6!50!black,fill=color6},
	{RYB7!50!black,fill=RYB7},
}

\pgfplotsset{
	select row/.style={
		x filter/.code={\ifnum\coordindex=#1\else\def\pgfmathresult{}\fi}
	}
}

\pgfplotsset{
	/pgfplots/xlabel near ticks/.style={
		/pgfplots/every axis x label/.style={
			at={(ticklabel cs:0.5,-2)},anchor=near ticklabel
		}
	},
	/pgfplots/ylabel near ticks/.style={
		/pgfplots/every axis y label/.style={
			at={(ticklabel cs:0.5,-5)},rotate=90,anchor=near ticklabel}
	}
}

\newcommand{\fqp}{{\footnotesize \textsf{FQP}}}
\newcommand{\newfqp}{{\footnotesize \textsf{Diba}}}
\newcommand{\scnoc}{{\footnotesize \textsf{Diba's NoC}}}
\newcommand{\noc}{{\footnotesize \textsf{NoC}}}

\newcommand{\glacier}{{Glacier}}

\newcommand{\gs}{{\footnotesize \textsf{GSwitch}}}
\newcommand{\gss}{{\footnotesize \textsf{GSwitches}}}
\newcommand{\gsl}{{\footnotesize \textsf{GSwitch-A}}}
\newcommand{\gsd}{{\footnotesize \textsf{GSwitch-B}}}

\newcommand{\ls}{{\footnotesize \textsf{LSwitch}}}
\newcommand{\lc}{{\footnotesize \textsf{Collector}}}
\newcommand{\pu}{{\footnotesize \textsf{PUnit}}}
\newcommand{\pus}{{\footnotesize \textsf{PUnits}}}
\newcommand{\ord}{{\footnotesize \texttt{ORDERS}}}

\newcommand{\cmj}{{\footnotesize \textsf{Circular-MJ}}}
\newcommand{\hbsj}{{\footnotesize \textsf{HB-SJ}}}
\newcommand{\opb}{{\footnotesize \textsf{OP-Block}}}
\newcommand{\topbricks}{{\footnotesize \textsf{Topology Bricks}}}
\newcommand{\topbrick}{{\footnotesize \textsf{Topology Brick}}}
\newcommand{\aggregation}{{\footnotesize \texttt{Aggregation}}}
\newcommand{\groupBy}{{\footnotesize \texttt{GroupBy}}}
\newcommand{\orderBy}{{\footnotesize \texttt{OrderBy}}}
\newcommand{\brick}[1]{{\footnotesize \textsf{Brick\_#1}}}

%% file: tables_funcs.tex
\pgfplotsset{
	compat=newest,
	select coords between index/.style 2 args={
		x filter/.code={
			\ifnum\coordindex<#1\def\pgfmathresult{}\fi
			\ifnum\coordindex>#2\def\pgfmathresult{}\fi
		}
	},
	/pgfplots/ybar interval legend/.style={
		/pgfplots/legend image code/.code={%
			\draw[##1,/tikz/.cd,yshift=-0.2em,bar interval width=0.7,bar interval shift=0.5]
			plot coordinates {(0cm,0.8em) (5pt,0.6em)};},
	},
	/pgfplots/ybar legend/.style={
		/pgfplots/legend image code/.code={%
			\draw[##1,/tikz/.cd,bar width=3pt,yshift=-0.2em,bar shift=0pt]
			plot coordinates {(2*\pgfplotbarwidth,0.6em)};},
	},
}

\pgfplotstableread{
	Name	x	min	avg	max	
	0.1		1	0.67	0.67	4.33
	0.01    2	2.95	3.95	2.05 
	0.001	3	2.0999999999999996	8.1	4.9	
	0.0001	4	2.84	8.84	7.16
}{\mytableS}

\pgfplotstableread{
	Name	x	min	avg	max	
	0.1		1	0.72	0.72	2.2800000000000002
	0.01    2	3.3899999999999997	4.39	1.6100000000000003 
	0.001	3	1.9699999999999998	5.97	1.0300000000000002
	0.0001	4	0.08999999999999986	6.09	0.9100000000000001
}{\mytableNS}

\pgfplotstableread{
	Name	x	mino	avgo	maxo	minoo	avgoo	maxoo	minooo	avgooo	maxooo	mint	avgt	maxt	mintt	avgtt	maxtt	minttt	avgttt	maxttt					
	1.0		1	2.9		25.9	4.1		4.5		83.5	3.5		9.8		234.8	10.2	1.5		9.5		2.5		2.7		26.7	2.3		2.3		84.3	2.7
	0.5		2	2.8		65.8	4.2		6.7		155.7	5.3		4.3		346.3	5.7	    2.6		23.6	2.4		6.2		69.2	5.8		5.2		157.2	4.8
	0.1		3	8.46	82.46	10.54	3.3		174.3	4.7		5.7		367.7	2.3     2.6		32.6	3.4		2.7		78.7	5.3		4.3		174.3	4.7
	0.01    4	1.7		95.7	1.3		1.7		192.7	1.3		1.6		384.6	1.4	    1.4		47.4	0.6		1.0		96.0	1.0		1.0		193.0	1.0
	0.001	5	0.3		97.3	0.7		0.1		194.1	0.9		0.1		386.1	0.9	    0.4		48.4	0.6		1.3		97.3	0.8		1.1		194.1	0.9
}{\mytableSJMSJ}

%% file: abstract.tex
\begin{abstract}
	\justify 
Stream processing acceleration is driven by the continuously increasing volume and velocity of data generated on the Web and the limitations of storage, computation, and power consumption. Hardware solutions provide better performance and power consumption, but they are hindered by the high research and development costs and the long time to market. In this work, we propose our re-configurable stream processor (Diba), a complete rethinking of a previously proposed customized and flexible query processor that targets real-time stream processing. Diba uses a unidirectional dataflow not dedicated to any specific type of query (operator) on streams, allowing a straightforward placement of processing components on a general data path that facilitates query mapping. In Diba, the concepts of the distribution network and processing components are implemented as two separate entities connected using generic interfaces. This approach allows the adoption of a versatile architecture for a family of queries rather than forcing a rigid chain of processing components to implement such queries. Our experimental evaluations of representative queries from TPC-H yielded processing times of 300, 1220, and 3520 milliseconds for data streams with scale factor sizes of one, four, and ten gigabytes, respectively.
\end{abstract}

%% file: introduction.tex
\section{Introduction}\label{sec:introduction}
Digital technology is growing at a rapid pace. This growth is fueled
not only by the number of connected devices but also by the ever
increasing volume and velocity of the data collected, bringing forth
many challenges and opportunities~\cite{Jagadish14_bigdata,
  Jin15_bigdata, Stephens15_bigdata, Bello16_bigdata,
  Hashem15_bigdata}.

A large part of the information in the digital world is transient and
its value diminishes over time~\cite{Turner14_bigdata}.

Its immediate exploitation has spurred the need for real-time stream
processing approaches and systems.

Another example is the Square Kilometer Array (SKA), a large radio
telescope project planned for construction in Australia, New Zealand,
and South Africa. It will have a total collection area of
approximately one square kilometer. SKA will collect in excess of an
exabyte of data per day~\cite{Kitchin13_ska_datasize}. Research is
already underway on how to handle this volume of data and how to
effectively detect binary pulsars in the data in real
time~\cite{Heerden14_ska}. This type of data processing requires
effective prepossessing to extract valuable features to
improve the efficiency of the data analytics pipeline.

These types of processing requirements, as well as a continuing shift
toward \emph{Big Data}, present an exciting opportunity to study the
interplay of software and hardware. This can help us understand the
advantages and disadvantages of the current hardware and software
co-design space for distributed systems, which must be fully explored
before resorting to specialized systems such as application-specific
integrated circuits (ASIC), field-programmable gate arrays (FPGAs),
and graphics processing units (GPUs).  Each hardware accelerator has a
unique performance profile with enormous potential to compete with
software solutions or complement them~\cite{PHILIPCHEN2014314,
  Istvan19, istvan_providing_2018, salamat_nascent_2021} in many novel
applications such as blockchain consensus~\cite{SitBI21} and graph
processing~\cite{ChenCTCHWC22}.

The adoption of hardware accelerators, particularly FPGAs, is gaining
momentum in industry and academia. Cloud providers like Amazon are
building new distributed infrastructures that offer
FPGAs\footnote{Xilinx UltraScale+ VU9P fabricated using a 16 nm
  process, with approximately 2.5 million logic elements and 6,800
  digital signal processing (DSP) engines.} connected to
general-purpose processors (CPUs) using a PCIe
connection~\cite{amazon_ec2_f1_fpga_2016}. Furthermore, the deployed
FPGAs share the same memory address space with CPUs, leading to new
applications of FPGAs using co-placement and co-processing
strategies. Microsoft's configurable
cloud~\cite{microsoft_configurable-cloud-acceleration_fpga16} also
uses a layer of FPGAs between network switches and servers to filter
and manipulate dataflows at line rate. Another prominent deployment
example is IBM's FPGA-based acceleration within the SuperVessel
OpenPOWER development cloud~\cite{ibm_supervessel_2016}. Google's
tensor processing unit (TPU), designed for distributed machine
learning workloads, is also gaining traction in its data centers; TPUs
are based on ASIC chips~\cite{google_tpu}.

Many solutions in stream processing have been developed in recent
years, such as~\cite{medusa, Arasu_steam16, Motwani03queryprocessing,
  Chandrasekaran_TelegraphCQ2003, Cranor2003_gigascope,
  Hammad_Nile2004, Zaharia_spark2010, ZhangHZH19BriskStream}, which
are primarily software-based approaches,
and~\cite{Mueller09_QueryCompilerFPGA, retefpgaToPSS, netezzaTwinFin,
  sparcm72015, Wu2014_q100, chen21_dac}, which benefit from a closer
hardware utilization. However, many of these approaches provide either
no support or limited support for hardware acceleration. The main
reason for this limitation is the complex and time-consuming process
of hardware design, development, and adoption, which even renders the
hardware solutions out of reach, further exacerbated by the constant
demand to evolve and expand the application
requirements~\cite{michell97_hwswcodesign}.

Existing proposals (e.g.,~\cite{Mueller09_QueryCompilerFPGA,
  Wu2014_q100, netezzaTwinFin, LuCWZ0Y22, KorolijaKKTMA22Farview,
  RoozkhoshHMPSD023, serot_implementing_2011}) aim to reduce the
complexity of hardware design by providing a library of building
blocks to synthesize query evaluation pipelines. However, the proposal
by Mueller et al.~\cite{Mueller09_QueryCompilerFPGA} is restricted to
providing a single query execution pipeline and does not allow the
synthesis of multiple query execution pipelines. Serot et
al.~\cite{serot_implementing_2011} propose CAPH, which is a DSL-based
approach for implementing stream processing applications on
FPGAs. Many other contemporary appraoches do not support the data
stream query processing model with sliding window
semantics~\cite{Wu2014_q100, netezzaTwinFin, LuCWZ0Y22,
  RoozkhoshHMPSD023}.

Thus, we advocate for a single system design and implementation
framework that supports the following requirements to allow for the
efficient and high-performant query processing over data streams:

\textbf{Efficient unidirectional on-chip dataflow}. Existing
approaches for data stream query processing either do not address
on-chip communication challenges
(e.g.,~\cite{sukhwani_hardwaresoftware_2015}) or use a complex
approach that requires additional buffering and bidirectional stream
flows for correct on-chip communication (e.g.,~\cite{najafi_fqp_vldb,
  najafi_fqp_icde}).

\textbf{Modular architecture and components}. System modularity
facilitates the reduction of design and implementation complexities of
hardware systems. A framework having modular components for
computation and communication can support rapid prototyping to meet
application demands. Many of the existing proposals for query
processing in hardware lack this aspect
(e.g.,~\cite{zhou_hardware-accelerated_2019,
  lasch_bandwidth-optimal_2022, ziener_onthefly_2012}).  Furthermore,
having a modular architecture simplifies synthesizing, deploying, and
executing multiple queries over data streams in hardware. Existing
modular approaches do not support running multiple queries
concurrently (e.g.,~\cite{Mueller09_QueryCompilerFPGA}).

\textbf{Online Reconfigurability}. Many of the existing stream
processing systems do not allow online reconfiguration of synthesized
queries (e.g.,~\cite{Mueller09_QueryCompilerFPGA}). While supporting
predetermined queries is useful in some application domains, having
the ability to reconfigure query processing during operation is highly
desirable to help applications adapt to changing workload
characteristics.

To satisfy the above requirements, we propose \newfqp{}, a
reconfigurable stream processing framework for designing and
implementing query processing engines over data streams in
hardware. \newfqp{} leverages a stream-aware network-on-chip design, a
comprehensive library of hardware components, and a simple interface
and methodology to customize and add components. \newfqp{} uses a
unidirectional communication flow to eliminate redundant buffering for
efficient communication. Moreover, the modular architecture of
\newfqp{} supports constructing stream query processing engines with
various degrees of flexibility that allow for fine-grained and
coarse-grained adjustments for specific applications. \newfqp{}
includes a library of programmable components that can be customized
for more efficient processing, communication, and workload
adaptation. Computational components include all necessary stream
query operators such as projection, selection, join, grouping, and
aggregation. Furthermore, \newfqp{} simplifies the deployment of
multiple queries that are executed concurrently in hardware and
supports the online reconfiguration of deployed queries. The
contributions of this paper can be summarized as follows:

\begin{enumerate}
    \item We introduce \newfqp{} and discuss its design in detail,
      including its unidirectional dataflow for enabling efficient
      on-chip communication;
    
    \item we introduce the modular design and architecture of each
      computational and communication component of \newfqp{};
    
    \item we provide an efficient component library that includes
      various components that serve as building blocks for
      instantiating different stream query processing engines;
    
    \item we introduce a new design for the hash-based stream join
      operator (\hbsj), which enables fast equijoin processing as well
      as for supporting \aggregation{} and \groupBy{} operators; and

    \item we evaluate representative queries from standard benchmarks
      (e.g., TPC-H) on a prototype implementation of \newfqp{}.
\end{enumerate}

The remainder of this paper is organized as
follows. Section~\ref{sec:related_work} describes related approaches.
Section~\ref{sec:overview} provides an overview of \newfqp{}'s
design. We then describe \newfqp{}'s modular architecture and
illustrate its benefits in
Section~\ref{sec:arch}. Section~\ref{sec:library} describes
\newfqp{}'s components in detail.  Finally, we present the main
results of our experiments in Section~\ref{sec:exp} using
industry-standard benchmarks, followed by concluding remarks in
Section~\ref{sec:conclusion}.

%% file: related_work.tex
\section{Related Work}\label{sec:related_work}
	
Due to its importance, continuous query processing over data streams
has received significant attention from the database research
community. The work related to \newfqp{} can be categorized as
follows.

\textbf{Software-based Stream Processing Systems.}  Early stream
processing approaches focused on centralized software-based data
stream processing engine designs (e.g.,~\cite{abadi_aurora_2003,
  Chandrasekaran03_TelegraphCQ, Arasu_steam16,
  Motwani03queryprocessing}). These approaches extend the relational
database model in various ways to efficiently support data stream
processing in software while running on commodity hardware. Unlike the
aforementioned work, \newfqp{} uses hardware acceleration to
accelerate the processing of relational queries over data streams. To
address the scalability and fault-tolerance issues with centralized
solutions, distributed stream processing engines have been developed
(e.g.,~\cite{balazinska_fault-tolerance_2008,
  toshniwal_stormtwitter_2014, carbone_apache_2015}).  More recently,
SCABBARD was introduced~\cite{theodorakis_scabbard_2021} which is a
single-node fault-tolerant stream processing engine. SCABBARD relies
on the concept of parallel persistence to scale up to high data stream
ingestion rates. \newfqp{} complements the ideas proposed by SCABBARD
and can result in better performance scalability by accelerating
workers' processing tasks. With Grizzly~\cite{grulich_grizzly_2020},
Grulich et al. propose an adaptive compilation framework to compile
queries for more efficient execution. Unlike \newfqp{}, Grizzly does
not use hardware accelerators such as FPGAs and relies on
parallelization techniques for NUMA-based CPU
architectures. LightSaber~\cite{theodorakis_lightsaber_2020} focuses
on processing window-based aggregation queries over streams by
leveraging modern CPU architectures.

\textbf{Application-specific Stream Processing Systems.}  While the
aforementioned proposals target general stream processing
applications, there have been some proposals that target specific
application domains. For instance, Gigascope~\cite{gigascope} is a
stream database for network applications in telecommunication
networks, and Tornado~\cite{mahmood_tornado_2015} supports distributed
query processing over spatio-textual data
streams. CPOD~\cite{tran_real-time_2020} performs distance-based
outlier detection. Maschi et al.~\cite{maschi_making_2020} leverage
hardware acceleration in Business Rule Management Systems (BRMS).

Comparatively, \newfqp{} complements all aforementioned approaches and
can be extended to support new application domains using its modular
component design framework. However, these extensions go beyond the
current scope of this paper.

\textbf{Hardware acceleration in Data Processing Systems.}  Moreover,
the idea of accelerating database operations using hardware is not
new, and several approaches exist in the literature.  Sun et
al.~\cite{sun_fpga-based_2020} accelerate compaction tasks in
log-structured storage
systems. FCAccel~\cite{watanabe_column-oriented_2019} use FPGAs to
accelerate column-oriented
databases. BionicDB~\cite{kim_bionicdb_2019} aims to accelerate OLTP
workloads using FPGAs.  Erylimaz et al.~\cite{eryilmaz_fpga_2021}
focus on studying the acceleration of aggregation operation in
database query processing.

ReProVide~\cite{g_sql_2020} uses FPGA to accelerate query processing
in a distributed/federated database system by co-processing part of
the query evaluation plan in hardware. ReProVide does not directly
support processing queries over data streams and does not use a
well-defined topology or modular building block for various query plan
operators. Sukwani et al.~\cite{sukhwani_hardwaresoftware_2015}
propose offloading certain operator processing to FPGAs for
acceleration. However, their system does not support joins or
\groupBy{} operators. Ziener et al.~\cite{ziener_onthefly_2012}
propose an FPGA-based partially re-configurable query processing
system that uses a module library. However, their module library
focuses on simple projection and selection queries. It does not
support complex queries that involve joins. Furthermore, their
approach lacks a modular architecture that can efficiently synthesize
multiple queries, and they focus on synthesizing a single query at a
time. Lasch et al.~\cite{lasch_bandwidth-optimal_2022} and Zhou et
al.~\cite{zhou_hardware-accelerated_2019} focus on improving the
performance of relational join operators in relational database query
processing. Neither of those approaches uses a modular system design
while accelerating join queries.

With respect to synthesizing queries over streams in hardware, Mueller
et al.~\cite{Mueller09_QueryCompilerFPGA} proposed \glacier{}, which
is a compositional query compiler that synthesizes queries on FPGAs
using a common component library. In \glacier{}, processing components
follow only a single query evaluation pipeline, making it more complex
to support synthesizing multiple queries.

More recently, Korolija et al. proposed offloading operators to
hardware accelerators with
Farview~\cite{KorolijaKKTMA22Farview}. Their approach does support
join computation acceleration while \newfqp{} additionally provides a
comprehensive framework for accelerated stream query processing
involving a multitude of operators.

SABER~\cite{koliousis_saber_2016} and
Finestream~\cite{zhang_finestream_2020} are platforms for executing
SQL queries over data streams on heterogeneous processing units of
CPUs and GPGPUs. In \newfqp{}, queries are executed on the synthesized
processing units.

In \newfqp{}, we support join and aggregation operations in the
context of hardware-based stream data processing. Furthermore, using
our modular design framework, \newfqp{}'s component library can be
extended to support more complex operations and processing pipelines
by designing new query processing building blocks.

\textbf{Heterogeneous Cloud Computing.}  More recent work accelerating
cloud-based workloads and serverless architectures
include~\cite{du_serverless_2022, fahmy_virtualized_2015,
  bacis_blastfunction_2020, werner_hardless_2022,
  istvan_providing_2018}.  Du et al.~\cite{du_serverless_2022}
introduce Molecule, which allows serverless applications to utilize
heterogeneous processing units (e.g., CPUs, DPUs, GPUs, and
FPGAs). Molecule achieves this by providing a unified abstraction
middleware that maintains the computational states of serverless
functions and facilitates their interaction with different processing
units. Towards a similar goal for serverless AI applications, Werner
and Shirmer~\cite{werner_hardless_2022} propose Hardless, which uses a
pre-configured Python-based runtime to run workloads on heterogeneous
hardware.

Bacis et al.~\cite{bacis_blastfunction_2020} propose the idea of
FPGA-as-a-service and focus on addressing the challenge of effectively
sharing FPGA resources across multiple tenants.  Istvan et
al.~\cite{istvan_providing_2018} develop FPGA-based resource-sharing
logic to support multi-tenanted key-value stores in clouds The work by
Fahmy et al.~\cite{fahmy_virtualized_2015} is among the earliest
proposals for providing cloud-based access to FPGAs.  Unlike
\newfqp{}, none of these approaches support processing queries over
streams. However, they provide interesting perspectives on future
research directions for \newfqp{}.

\textbf{Commercial Database Acceleration Systems.}  While \newfqp{}
focuses on leveraging hardware acceleration for processing queries
over data streams, there have been many proposals for accelerating
query processing in databases in general. For instance, Wu et
al. ~\cite{Wu2014_q100} proposed a domain-specific database processor
named Q100 to handle database applications efficiently. Q100 contains
a heterogeneous collection of fixed-function ASIC tiles, each of which
implements a well-known relational operator, such as a join or
sort. Furthermore, another commercially available product is IBM
Netezza ~\cite{netezzaTwinFin}, which exploits parameterizable
circuits to offload query computation and accelerate data handling and
processing.

In contrast to Q100 and Netezza, which are built for database
operations, \newfqp\ is designed for data stream processing, which
fundamentally changes the design requirements of the hardware
architecture. Moreover, \newfqp\ is designed to support various query
types in stream processing. Using \newfqp{}, components can be
rearranged to improve the utilization of almost any type of processing
component. The NoC designed for \newfqp{} emphasizes the single-input
and single-output model to improve the communication efficiency among
components.

\textbf{Comparison to our Previous Work.}  The most relevant work to
\newfqp{} is our previous work referred to as
\fqp{}~\cite{najafi_fqp_vldb,najafi_fqp_icde}; many lessons learned
from building \fqp{} influenced the design of \newfqp{}.

Compared to \fqp{}, \newfqp{} differs in four major design
choices. First, the data flow of the physical query execution
pipeline: \fqp{} uses a bidirectional dataflow, which allows \fqp{}
the parallel processing of join operations involving multiple data
streams, as shown in Figure~\ref{fig:biflow}. This design requires
complex mappings to place operators and route intermediate results
from the output of one operator to the input of another.

Unlike \fqp{}, in \newfqp{}'s architecture, the data flow is
unidirectional and is not dedicated to any specific type of
stream. This architecture is made possible by the introduction of join
parallelization to the stream using a single data
path~\cite{najafi_splitjoin}. As a result of this architecture, the
processing components can be placed one after another instead of the
complex arrangement in \fqp{}.

The second choice is to design for modularity in the \newfqp{}
architecture, where we separate the distribution network and data
processing components. We use interfaces to connect the distribution
network and processing components. This design allows adding
processing components where needed in the distribution network, which
routes corresponding streams to each component using defined
instruction sets. This modularity enables a simplified placement of
queries. In contrast, \fqp{} uses a complex algorithm for placing
query operators on a rigid chain of processing components.

Third, in terms of hardware, the number of ports a component exposes
closely relates to its cost, complexity, and efficiency.  In
\newfqp{}'s architecture, the number of ports is greatly reduced, and
it is based on a consume-and-produce model. Processing and
transmission components have one input port and one output port. In
contrast, \fqp{}'s components have five ports (two input and three
output.) \newfqp{} effectively eliminates the need for large and
complex switching and control circuits responsible for consistently
receiving, processing, and transmitting streams and results from
multiple ports when compared to \fqp{}.

Finally, in \newfqp, tuples for distinct streams involved in the join
operation arrive in order. In contrast, in \fqp, there are two
separate paths leading to out-of-order arrivals, which require complex
buffering, and control circuits are required to ensure the consistency
of the results.

%% file: overview.tex
\section{Diba Overview}
\label{sec:overview}

\begin{figure}
	\centering
	\includegraphics[width=0.45\linewidth]{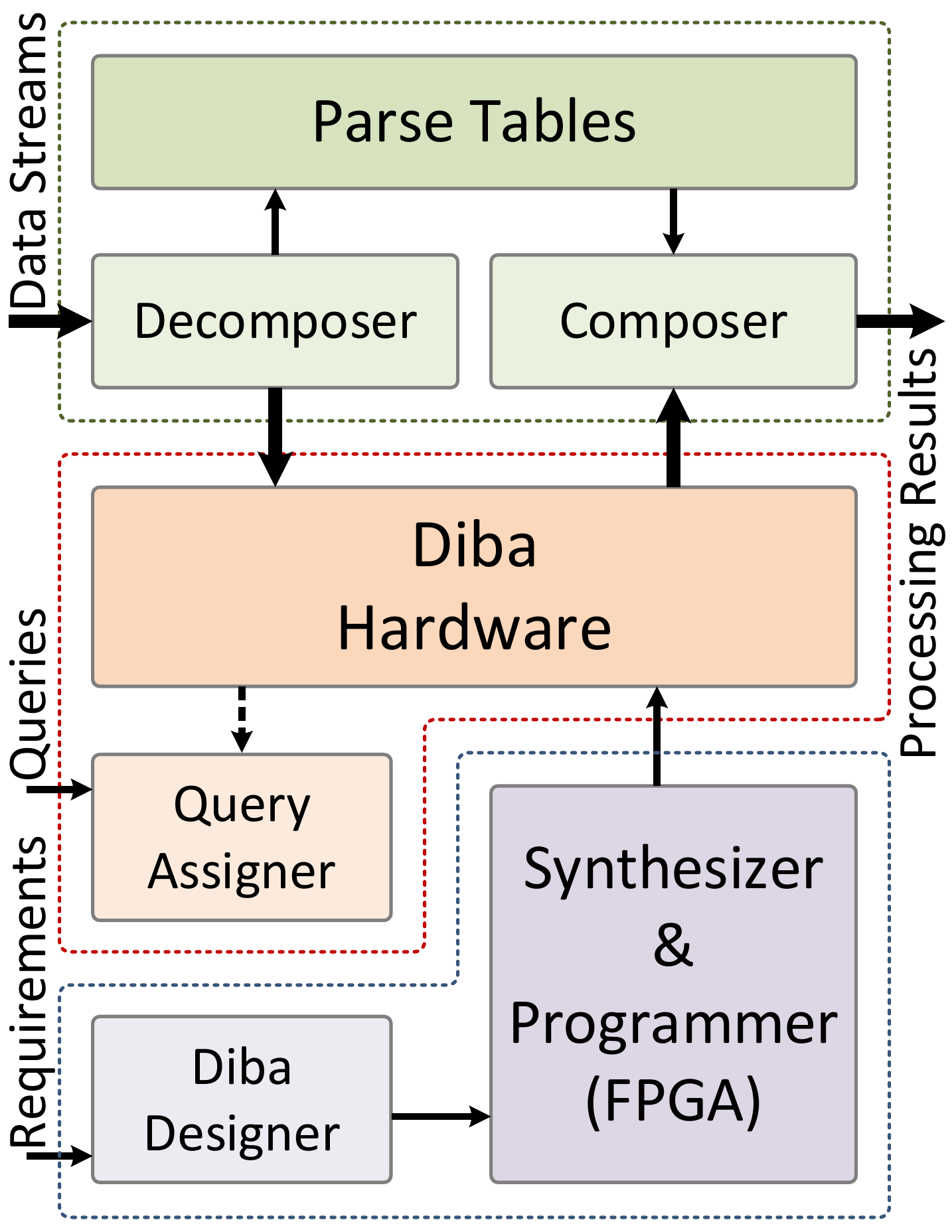}
	\caption{\newfqp{} framework architecture}
	\label{fig:fqpv2completesystem}
\end{figure}

The stream processing system that integrates \newfqp{} has the
architecture as shown in Figure~\ref{fig:fqpv2completesystem}. A data
stream is conceptually defined as an infinite sequence of data tuples
that continuously arrive from different data sources over which
queries are evaluated. At the top, incoming data streams are processed
by a \textit{Decomposer}, which extracts the parts of the incoming
tuples that are needed for query processing and forwards them to
\newfqp{}'s query processing engine implemented in hardware and keeps
the parts that are not used by the processing engine in \textit{Parse
  Tables}. The information stored in Parse Tables is later used by the
\textit{Composer} to reconstruct the query result streams.

When working with specialized hardware, filtering unimportant (for the
registered queries) parts of data is crucial to delivering a practical
and efficient system. These parts need to be reattached to the
resulting tuples. For example, suppose that we have an age attribute
in our tuples that is not used in the given query; however, we need
this attribute in the resulting tuples. To avoid using valuable
hardware resources to transfer this attribute, we cut it from the
input tuples and reattach it to the corresponding resulting output
tuples.

The proposed design of \newfqp{} can accept new query registrations or
updates to already registered queries.  To achieve this flexibility,
the \textit{Query Assigner} takes query specifications and deploys
them on the processor engine. The Query Assigner maps new query
specifications to the available processing blocks and generates the
corresponding instructions to program the query. These instructions
are fed into the \newfqp\ hardware through the \textit{Data Streams}
input port.

Finally, the \textit{Diba Designer} allows designing custom topologies
using a graphical user interface and built-in component
libraries. This approach abstracts away the complexity and the details
of hardware design and development. Following topology design, it is
first synthesized and then programmed into the FPGA.

\subsection{Unidirectional Dataflow}

\begin{figure}[t]
	\centering
	\begin{subfigure}[t]{0.4\linewidth}
		\centering
		\includegraphics[width=0.9\linewidth]{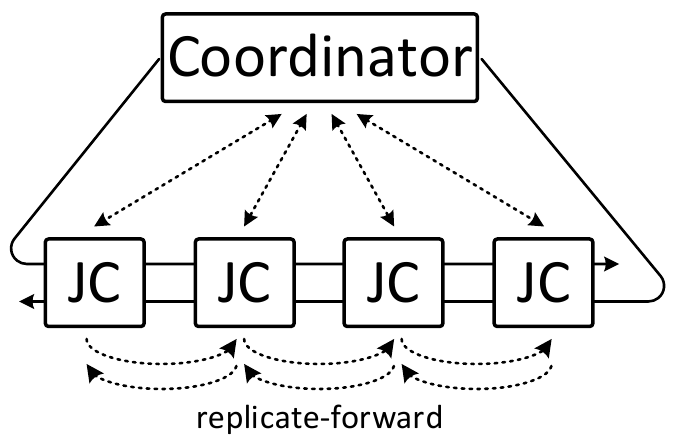}
		\caption{Bidirectional (\fqp{})}
		\label{fig:biflow}
	\end{subfigure}
	\hfil
	\begin{subfigure}[t]{0.4\linewidth}
		\centering
		\includegraphics[width=0.78\linewidth]{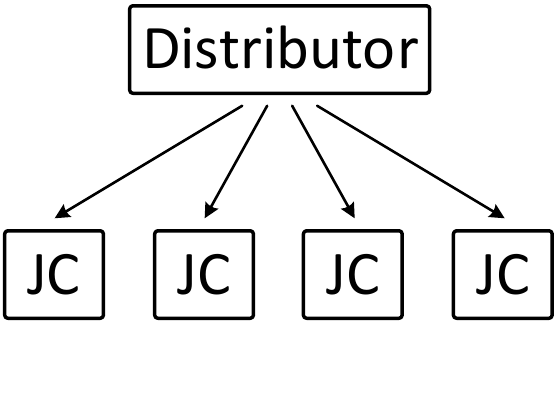}
		\caption{Unidirectional (\newfqp{})}
		\label{fig:uniflow}
	\end{subfigure}
	\caption{Stream join data-flow models}
	\label{fig:dataflow}
\end{figure}

In this section, we discuss the details of bidirectional data flow in
the context of stream-join processing. \fqp\ utilizes bidirectional
data-flows (Figure~\ref{fig:biflow}), which is inspired by the
handshake join algorithm~\cite{Teubner11_Handshake}. The bidirectional
data flow supports stream-join parallelization. It uses
\textit{programm\-able bridges} to connect the chains to one another
and to the input and output of the processor. Although this design
enables FQP~\cite{najafi_fqp_vldb,najafi_fqp_icde,najafi_fqp_sigmod}
to accelerate stream join processing by spreading the query over
multiple join cores (JCs), realized using \underline{o}nline
\underline{p}rogrammable \underline{blocks} (\opb). Unfortunately, in
practice, this approach introduces new challenges. First, it requires
that processing blocks have five ports. In hardware solutions, the
number of input/output ports is related to the cost, complexity, and
efficiency of a component. Second, the mapping of operators to the
processing blocks is challenging. As illustrated in
Figure~\ref{fig:biflow}, the chain of processing blocks is connected
to the remaining components at its two ends. This layout requires a
linear arrangement of all operators in a query. Although this
limitation is mitigated by the available bypass unit inside the
blocks, it still significantly limits the number of operators that
work in parallel. Third, the utilization of more than two streams is
difficult, where the bidirectional data flow favors two streams by
devoting a data path to each. \fqp\ partially addresses this issue by
sharing each data path among more streams but with additional
controlling overhead to bypass blocks (operators) that are not used in
the processing of specific streams.

To overcome the aforementioned challenges with bidirectional data flow
and reduce the complexity, \newfqp{} adopts a unidirectional data-flow
approach. Our new communication architecture extends ideas from Najafi
et al.~\cite{najafi_splitjoin} while providing the performance
benefits of stream-join parallelization and the usability benefits of
mapping query operators to processing components.

\subsection{Adopting a System-on-Chip approach for Diba}
A processing engine in hardware that integrates all or most of the
necessary components to perform a task is referred to as a
system-on-a-chip (SoC). SoCs have the best-in-class performance and
efficiency because they bring these components close to one another,
removing the most unnecessary inter-component interfaces and their
added latency. SoCs are particularly interesting in data stream
handling owing to the high velocity and volume of streams because
every added latency to the processing path superlinearly increases the
cost of the system.

\subsubsection{Background}

Each SoC solution benefits from a particular type of communication
sub-system to transfer data between various components, such as a
processor, a coprocessor, and a DMA\footnote{Direct memory access.}
controller, and peripherals. SoCs commonly use a \textit{bus} (e.g.,
the ARM AMBA\footnote{Advanced microcontroller bus architecture.})
system that acts as a communication subsystem. A bus consists of an
access controller and a set of wirings that connect all internal
components of an SoC.

Despite the simplicity of the conventional bus architecture, it fails
to scale when the number of connected components increases. Thus, we
are forced to rely on distributed architectures, i.e., network-on-chip
(NoC). NoC solutions have their own challenges and opportunities. They
provide a substantial bandwidth advantage over bus subsystems, but at
the expense of extra controlling logic (in terms of both hardware and
software) to ensure consistency of communication.

\subsubsection{Main Drawback}

Typical NoCs are designed for general use cases to handle a wide range
of workloads. However, this generality is not needed or can at least
be reduced in stream processing in favor of simpler and more efficient
communication circuitry, resulting in better efficiency and
performance characteristics.

A data stream is a (potentially unbounded) sequence of tuples that
flow one after another. This flow-based property of data streams
motivates a processing architecture that avoids circulating data
between the internal components. For example, consider a processing
unit performing aggregations that uses another unit to perform
filtering on the intermediate result data. The unfiltered data is sent
to the filtering unit before the filtered data is sent back (e.g., to
perform additional computations). Clearly, the above approach leads to
tuple circulation, which causes major issues in real-time
processing. A circular path in a digital system, constantly receiving
new data, inherently adds the possibility of deadlock that
necessitates complex buffering and communication circuitry to reduce
the likelihood of congestion.

\section{Diba's Modular Architecture}\label{sec:arch}

\begin{figure}
	\centering
	\includegraphics[width=0.8\linewidth]{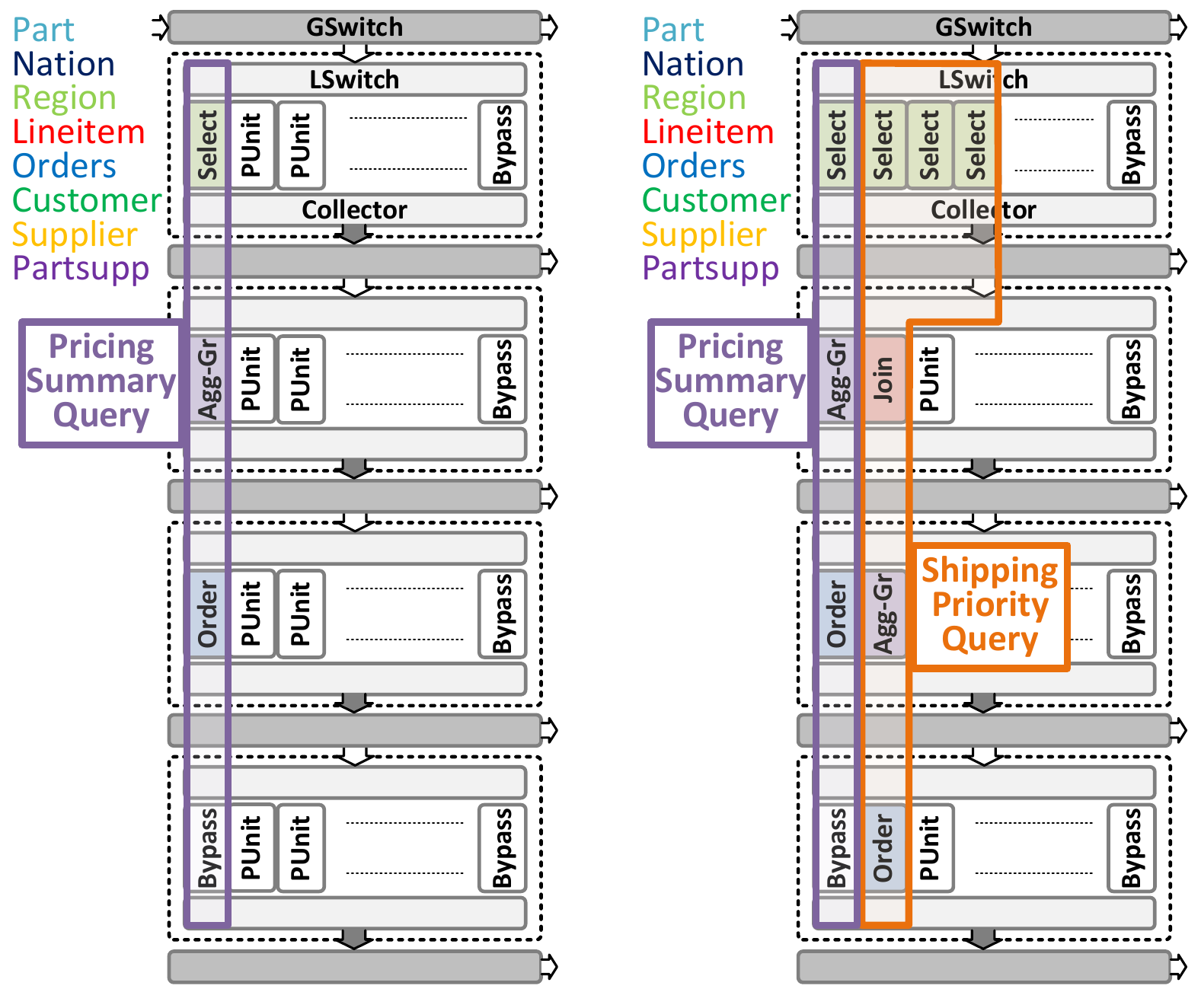}
	\caption{\newfqp{} modular implementation}
	\label{fig:custom_noc}
\end{figure}

This section discusses the modular architecture of \newfqp{}. In
\newfqp{}, we use custom-built stream processing components with a
network-on-chip (\noc{}) to leverage the flow-based property of
streams.

Figure~\ref{fig:custom_noc} presents a logical model of a simplified
implementation of \scnoc{}, which aims to illustrate \newfqp{}'s
features such as modularity, reconfigurability, and efficiency. In the
figure, we show some of its main components, such as \gsl\ and \ls\,
that perform the main data routing tasks. After implementing a
topology of \scnoc\ that is reconfigurable based on the application
requirements, the routing instructions are fed to \scnoc\ to program
{\gsl}s and {\ls}s. Subsequently, streams of data are routed and
brought to their corresponding processing units (\pus{}) by the
\gsl\ and \ls\ components.

On the left side of Figure~\ref{fig:custom_noc}, we show the instant
in time when the query called \emph{Pricing Summary Query (PS)} is
deployed. On the right side of the figure, we show an example
deployment of an additional second query called \emph{Shipping
  Priority Query (SP)}. The SP query can be deployed at the same
instant in time as the PS query or later, which illustrates the
reconfigurability of our design. Due to the modular architecture of
\newfqp{}, users can easily reason about the deployment and reuse of
building blocks (e.g., \topbricks). More queries can be deployed
depending on the implementation's specifications to increase the
overall utilization of the hardware. Furthermore, \newfqp{}'s
architecture allows cost-effective configurations that can share
intermediate results.

\begin{figure}
	\centering
	\includegraphics[width=0.5\linewidth]{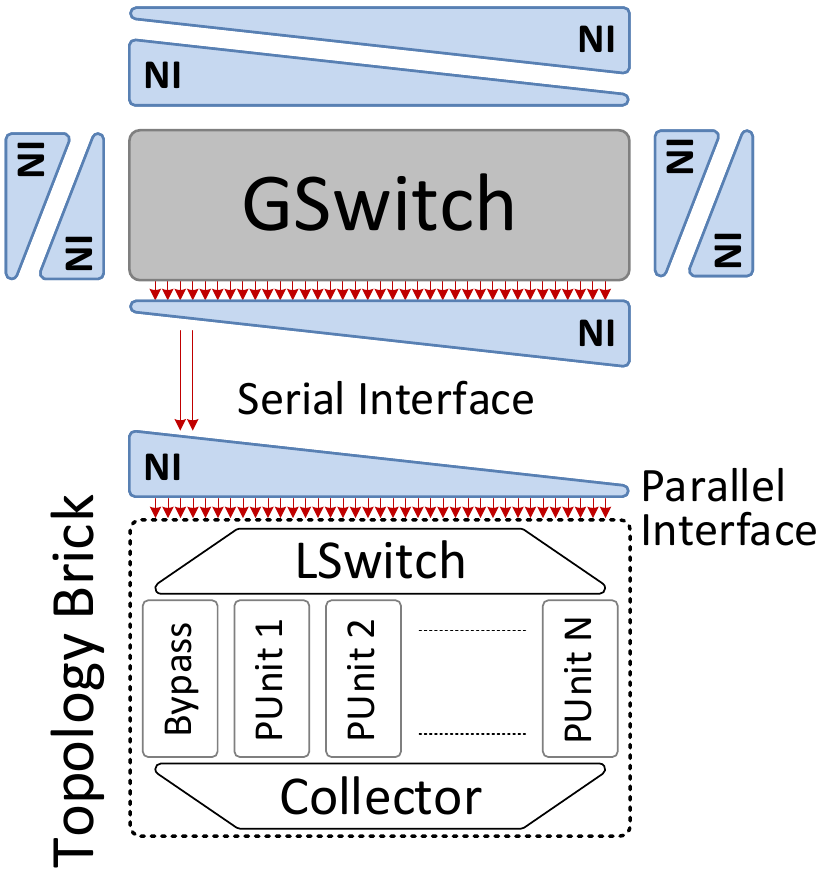}
	\caption{Modular communication interface}
	\label{fig:n_interface}	
\end{figure}

\begin{figure*}
	\centering
	\begin{subfigure}[t]{0.3\linewidth}
		\centering
		\includegraphics[width=1.0\linewidth]{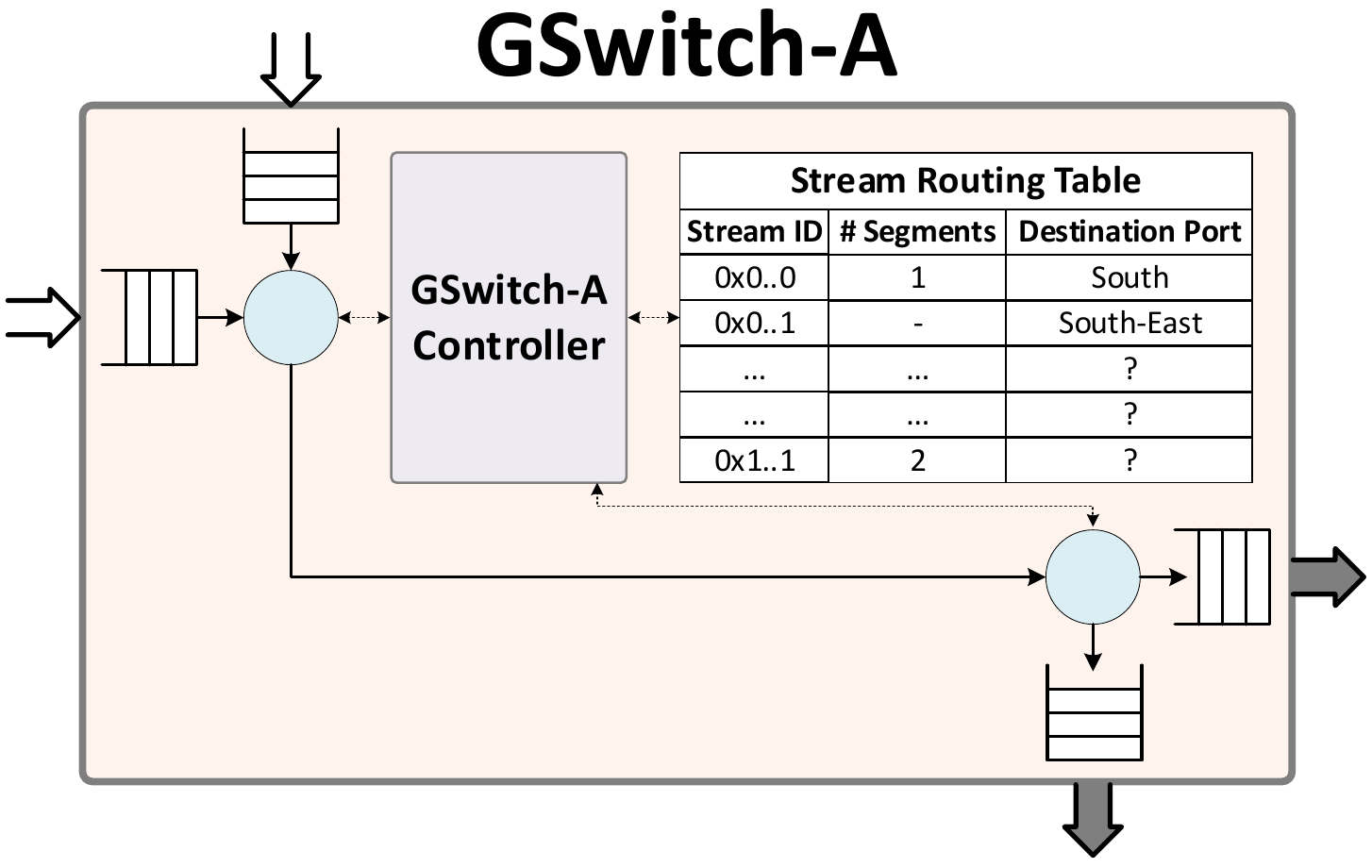}
		\caption{} 
		\label{fig:gswitchl}
	\end{subfigure}%
	\hfil
	\begin{subfigure}[t]{0.3\linewidth}
		\centering
		\includegraphics[width=0.9\linewidth]{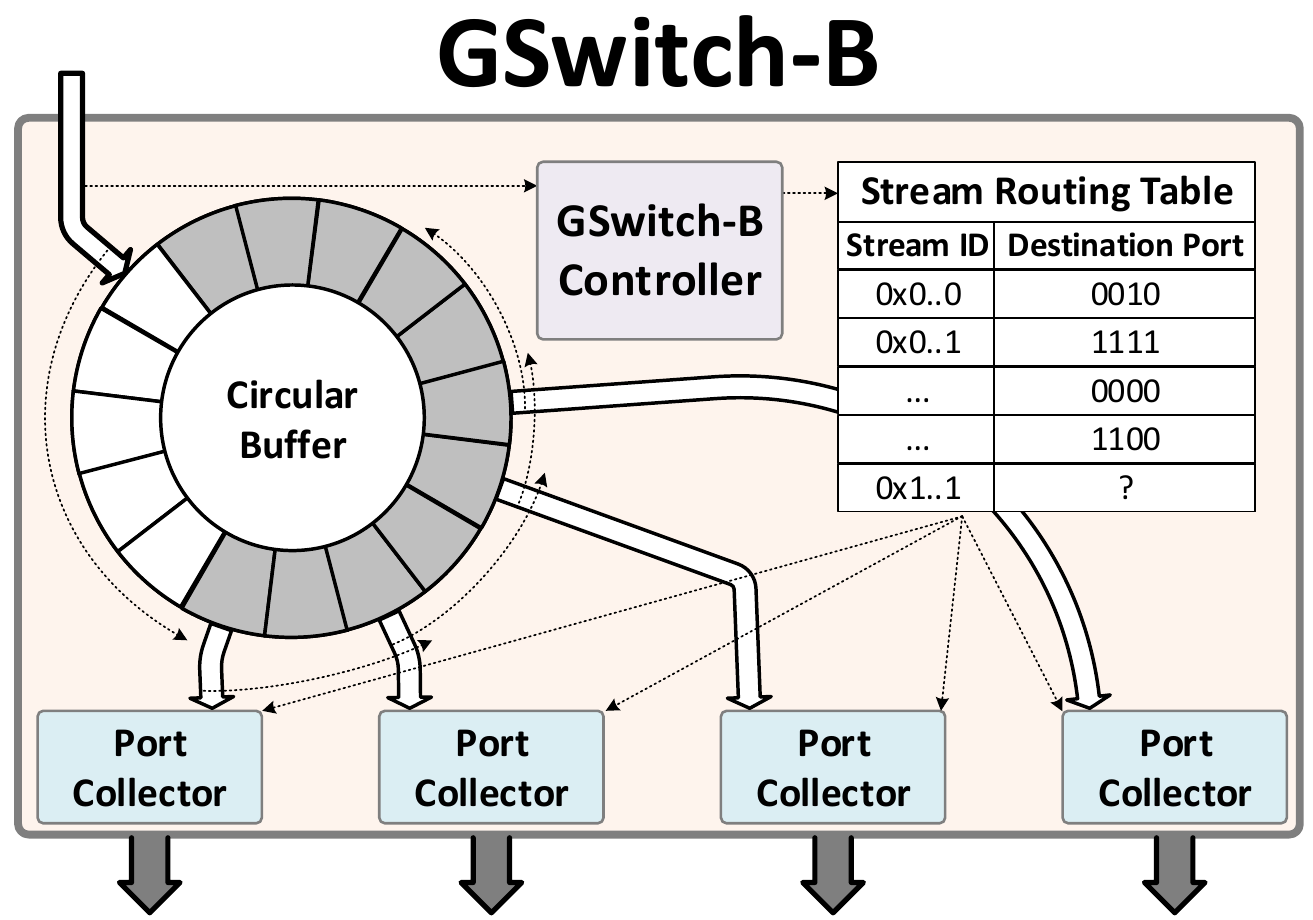}
		\caption{} 
		\label{fig:gswitchd}	
	\end{subfigure}
	\hfil
	\begin{subfigure}[t]{0.3\linewidth}
		\centering
		\includegraphics[width=0.9\linewidth]{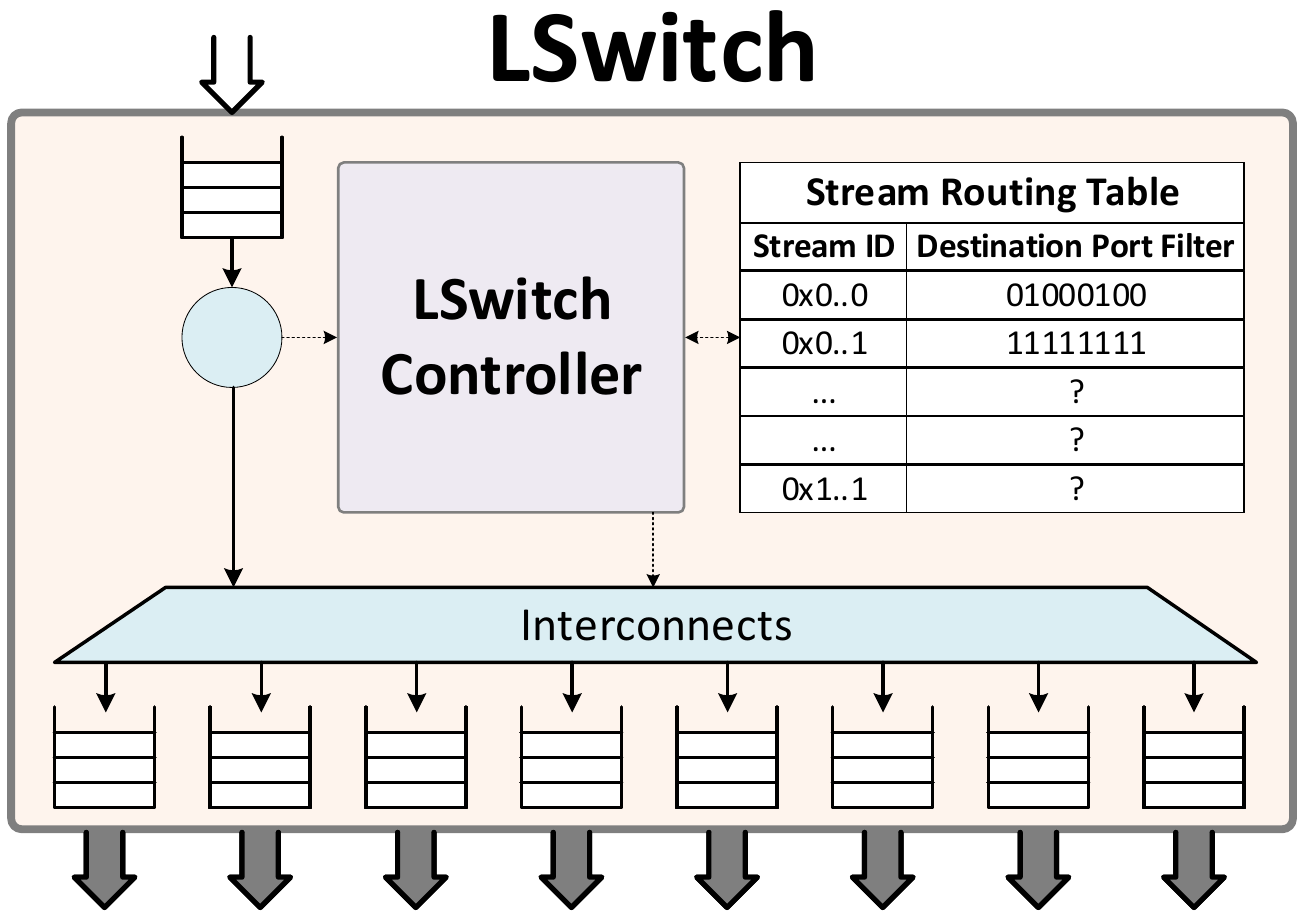}
		\caption{} 
		\label{fig:lswitch}
	\end{subfigure}
	\caption{\scnoc\ communication block}
	\vspace{-4mm}
\end{figure*}

\subsection{Topology Brick}

The processing components of \newfqp\ are packed, in addition to the
local data distribution and collection components, into specific sets
that we refer to as \topbricks{}, shown in the lower part of
Figure~\ref{fig:n_interface}. How the \topbricks{} are connected
defines the topology of a \newfqp\ instance that characterizes the
capabilities and properties of the system. Each \topbrick{} contains
two \newfqp{} \noc{} components of \ls\ and \lc, in addition to a set
of data handling and processing components specified by the
\textit{Bypass} and {\footnotesize \textsf{\pu{} 1..N}} labels in
Figure~\ref{fig:n_interface} described next.

\subsection{PUnits}

The \pus{} follow a straightforward protocol for receiving
data/instructions from an input port and pushing the resulting data
from their output port. They can implement any type of execution
engine, ranging from simple filtering units to highly specialized
hardware or even general-purpose processors (i.e., ARM cores). In
addition, \pus{} can have their own dedicated memory or use a shared
memory hierarchy with other \pus{} through a dedicated port based on
the application requirements.

\subsection{GSwitch}

\scnoc\ is constructed from programmable (\underline{g}lobal)
\underline{switch}es that we refer to as \gs. We propose two
architectures for the \gs{}, specified by -A and -B.

{\gss} are responsible for bringing data to a specific set of
\pus{}. The main properties of the design of a \gs\ are 1) data
pipelining to mitigate processing stalls, and 2) the number of flows
to be routed, or more precisely, the number of input ports
(\textit{x}) routed to the number of output ports (\textit{y}).

{\gss} identify data streams by an identifier and process them as
segments of tuples. Segments are chunks of data\footnote{The size of
  each chunk is defined statically (e.g., 64 bits) in our hardware
  specifications.}. Thus, tuples of a stream can consist of one or
more segments.  In this section, we give an overview of the design and
implementation of the two architectures. However, in our evaluation,
we use \gsl{} because there is no need to route tuples to more than
two collectors.

\subsubsection{\gsl}

\gsl\ is a variant of \gs\ with two input (\textit{x=2}) and two
output (\textit{y=2}) ports, as shown in
Figure~\ref{fig:gswitchl}. This unit has a \textit{stream routing
  table} that indicates the output port of each stream. The routing
information in this table is programmed using its corresponding
instructions. Each row in this table belongs to a stream and contains
three fields: 1) \textit{Stream ID}; 2) \textit{Number of Segments};
and 3) a \textit{Destination Port}, which can be one or both output
ports. The \textit{\gs\ controller} monitors incoming tuples from both
input ports. When there are instructions related to this unit, it
updates the table based on the tuples and then pushes them to the
output ports to reach their destination blocks.

The input and output buffers reduce stalls in the data flow because
they allow for the insertion of new data even when the processing
blocks in the outputs are busy. Therefore, these buffers improve
pipelining parallelism. In other words, the input and output buffers
allow the \textit{GSwitch controller} to route currently available
data ahead of time without waiting for the processing blocks in the
output.

\subsubsection{\gsd}

We refer to a \gs\ design customized for distribution as \gsd, which
is preferable when our application requires large data broadcasts,
i.e., at the input of \scnoc\ where streams are distributed according
to their selected processing paths. The abstract architecture of the
\gsd\ (\textit{x=1, y=bounded n}) is shown in
Figure~\ref{fig:gswitchd}. The design consists of a \textit{stream
  routing table} that preserves stream IDs with their corresponding
masks. There is a one-to-one assignment between each bit in the mask
field and its corresponding output port, and each port's bit
determines whether a stream must be sent out to that port. The masking
information is programmed through \scnoc{} instruction used by the
\textit{\gsd\ controller}.

In hardware, elements of internal memory are limited and expensive
resources. An interesting design aspect of \gsd\ is the use of a
shared circular buffer for all ports to reduce memory usage. Using a
first-in, first-out (FIFO) policy, each \textit{port collector}
benefits from the larger buffer, which adds more elasticity to the
processing data path. Using this shared buffer, we reduce the chance
of resource starvation in some \gsd\ output ports while fully loading
the processing units in the other output ports, which improves
processing efficiency and performance.

Compared with \gsl, which has two input and two output ports, \gsd\ is
more suited to large fan-outs, where a single input is fed to many
other units.

\subsection{\ls}

Each \gsl\ is connected to the processing and data handling blocks
through a local switch referred to as \ls. Figure~\ref{fig:lswitch}
illustrates the internal building blocks of \ls. This unit utilizes a
programmable $1-to-N$ switch to distribute tuples and instructions to
blocks connected to its output ports. The \textit{stream routing
  table} in \ls\ has two fields: 1) \textit{Stream ID} and 2)
\textit{Destination Port Filter}. \ls\ uses a masking technique to
specify the port(s) of egress for each stream.  For this reason, the
destination port filter uses a one-bit mask per port, which enables
data transmission over that port in the same manner as in \gsd. The
\textit{\ls\ controller} fills this table as it receives instructions
from the \ls\ input port. If an instruction does not belong to the
given \ls, this controller broadcasts it to all its output ports.

\ls\ is designed to support up to $N$ blocks in its output, where the
value of $N$ is determined based on the application
requirements. \textit{Interconnect} in Figure~\ref{fig:lswitch} is the
main component affected by the size of $N$. Using a small interconnect
can lead to the under-utilization of \scnoc\ because the number of
processing components (limited by $N$) is not sufficient to fully
utilize the bandwidth provided by components of the communication
network. Further, aiming for a large interconnect can negatively
affect the frequency of the working clock\footnote{A major parameter
  of the hardware that has a direct relation to processing
  performance.} of \scnoc, especially when parallel processing on a
large number of processing blocks is needed. One can exploit multiple
\gsl\ units to instantiate multiple \ls\ units, each with an
appropriate interconnect size.

\subsection{Collector}
Following the distribution of the input tuples to \pus{} and other
data handling blocks by \ls, the \lc\ (Figure~\ref{fig:n_interface})
gathers the resulting tuples from them. \lc\ can use a hierarchical or
linear architecture to gather the data, each with advantages and
disadvantages. The former architecture is suitable for many processing
blocks as the collection task can be performed in parallel for all
\pus{}. The latter architecture favors {\small \textsf{topology
    bricks}} with fewer blocks because it gathers the resulting tuples
one after another, which leads to poor scalability in relation to the
number of blocks; however, it requires fewer resources to operate.

If a processing block has resulting tuples that are wider than the
input, they are broken into smaller (as wide as the input)
segments. \lc\ must ensure that it gathers all segments of a tuple
from a block before starting the collection from another block. This
is done by assigning an unused (NULL) stream ID to all segments of a
tuple except the first one, which carries its origin stream
ID. Whenever \lc\ starts the collection task from a block, it
continues to gather all segments of a tuple one after another until it
observes a segment with an existing stream ID. It then starts
collecting a tuple from another block based on the defined priorities.

\subsection{Discussion} 
This section discusses three critical design choices we made for
\newfqp{}.

\subsubsection{Feedback Path}
A significant decision that heavily influences \scnoc\ architecture is
to limit internal (interblock) communication to unidirectional rather
than bidirectional communication. Although this eliminates the
possibility of returning data to a former processing block, it leads
to a data path suitable for real-time processing by forbidding
inefficient query mappings. In other words, the design of
\scnoc\ targets a paradigm where the processing is in a streaming
fashion, a flow. It is possible to add a feedback paths to a
\scnoc\ architecture by using a branch component
(Figure~\ref{fig:custom_noc}) instead of a \pu{} and feeding the
output port to one of the open entrances of the
\scnoc\ architecture. However, this approach is not recommended due to
additional complexities, such as race conditions. If such feedback is
necessary for processing a query, our approach is to implement the
entire feedback part (including its feedback path) in a single \pu{}.

\subsubsection{Instruction Set}

\begin{figure}
	\centering
	\includegraphics[width=0.95\linewidth]{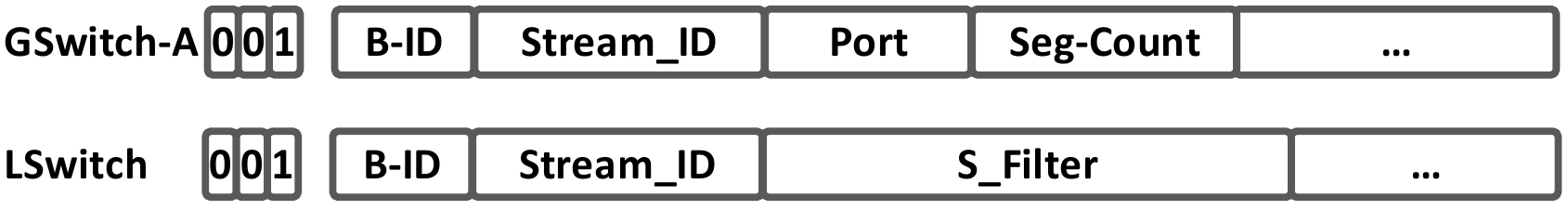}
	\caption{\gsl\ and \ls\ instruction sets}
	\label{fig:switchinst}
\end{figure}

As mentioned in the foregoing, the challenge in the design of
\scnoc\ is to maintain its architecture as simple and lightweight as
possible while providing sufficient routing flexibility to bring
streams to their corresponding processing blocks. In this regard, the
\gsl\ and \ls\ units are programmed using unique instructions, as
shown in Figure~\ref{fig:switchinst}.

At the beginning of each tuple, there are a few bits, e.g., two or
more bits, depending on the maximum number of supported streams, that
specify the ID of each stream and the type of data (instruction or
tuple). For example, a value of $1$ in the stream ID field
(Figure~\ref{fig:switchinst}) defines the given data as a network
instruction, whereas $0$ defines the data as a processing block
instruction. Other values are used to specify data stream IDs.

The instruction for \gsl\ consists of the following: 1) \textit{B-ID}
that defines the ID of the target block for this instruction, 2)
\textit{stream-ID} that specifies the stream in which this instruction
carries its routing information, 3) \textit{port} that determines the
mask of the egress port for the specified stream, and 4)
\textit{seg-count} that indicates the number of segments for each
tuple for the specified stream. The remaining fields are undefined and
can be utilized in future extensions. For example, data with fields
$``1:23:4:south:3:..."$ programs a \gsl\ with a block ID of $23$ to
route three segment-size tuples of a stream with ID $4$ to its $south$
port.

Similarly, the \ls\ instruction contains the \textit{B-ID} and
\textit{stream-ID} fields, but it uses \textit{an S-filter} (as shown
in Figure~6) to specify the egress port(s).

\subsubsection{Network Interface}

An inherent problem in custom hardware is its sensitivity to data
size. For example, when we design and build hardware for 64-bit-wide
data, this hardware cannot process wider data without considering them
in the initial design. The communication network in \scnoc{} is
designed to accept wider data units by breaking the data into smaller
segments. However, resource over-provisioning results in a significant
increase in cost as the width of the data increases, which renders it
impractical.

For a sizeable practical solution, we propose using parallel-to-serial
and serial-to-parallel network interfaces (NIs) to avoid the
additional wiring between internal
blocks. Figure~\ref{fig:n_interface} illustrates these interfaces in
triangle pairs, where NIs convert the parallel data into serial while
the other performs the reverse operation. The optimum number of serial
lines between each pair of NIs can be determined based on the
application and the technical specifications. For example, when the
processing effort and time in a \pu{} are significant, the results
come one after another in large intervals, which allows the use of
fewer serial lines. Having more transfers necessitates the use of more
serial lines. When full bandwidth is required, it is advisable to use
parallel communication rather than serial communication.

\section{Diba's Component Library}\label{sec:library}

In this section, we describe the various components available in
\newfqp{}'s component library.

\subsection{OP-Block}\label{sec:op_block}

\begin{figure}[t]
	\centering
	\begin{subfigure}[t]{0.39\linewidth}
		\centering
		\includegraphics[width=1\linewidth]{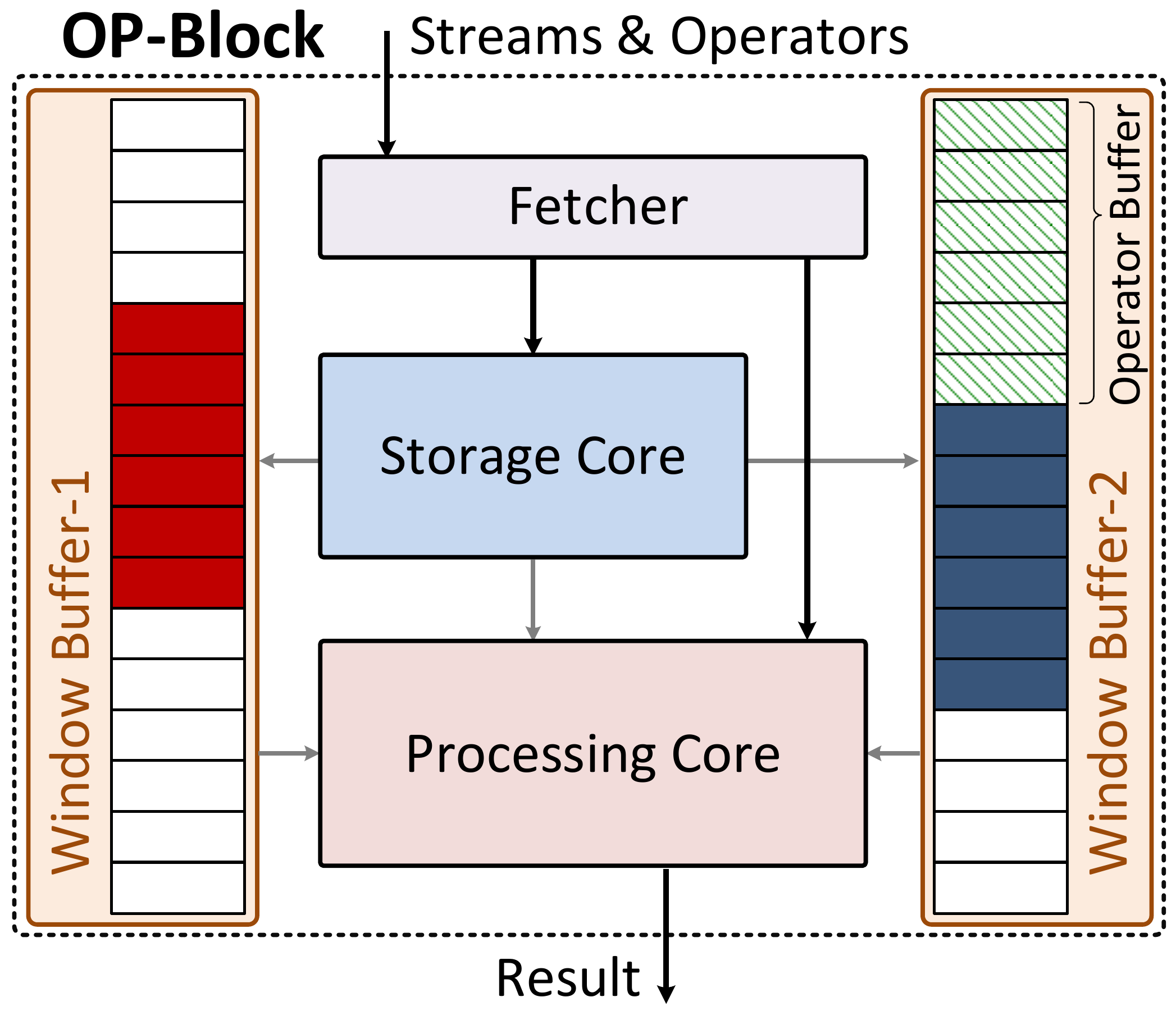}
		\vspace{-5mm}		
		\caption{Successor (in \newfqp{})}
		\label{fig:new_opblock}	
	\end{subfigure}
	\hfil
	\begin{subfigure}[t]{0.59\linewidth}
		\centering
		\includegraphics[width=1\linewidth]{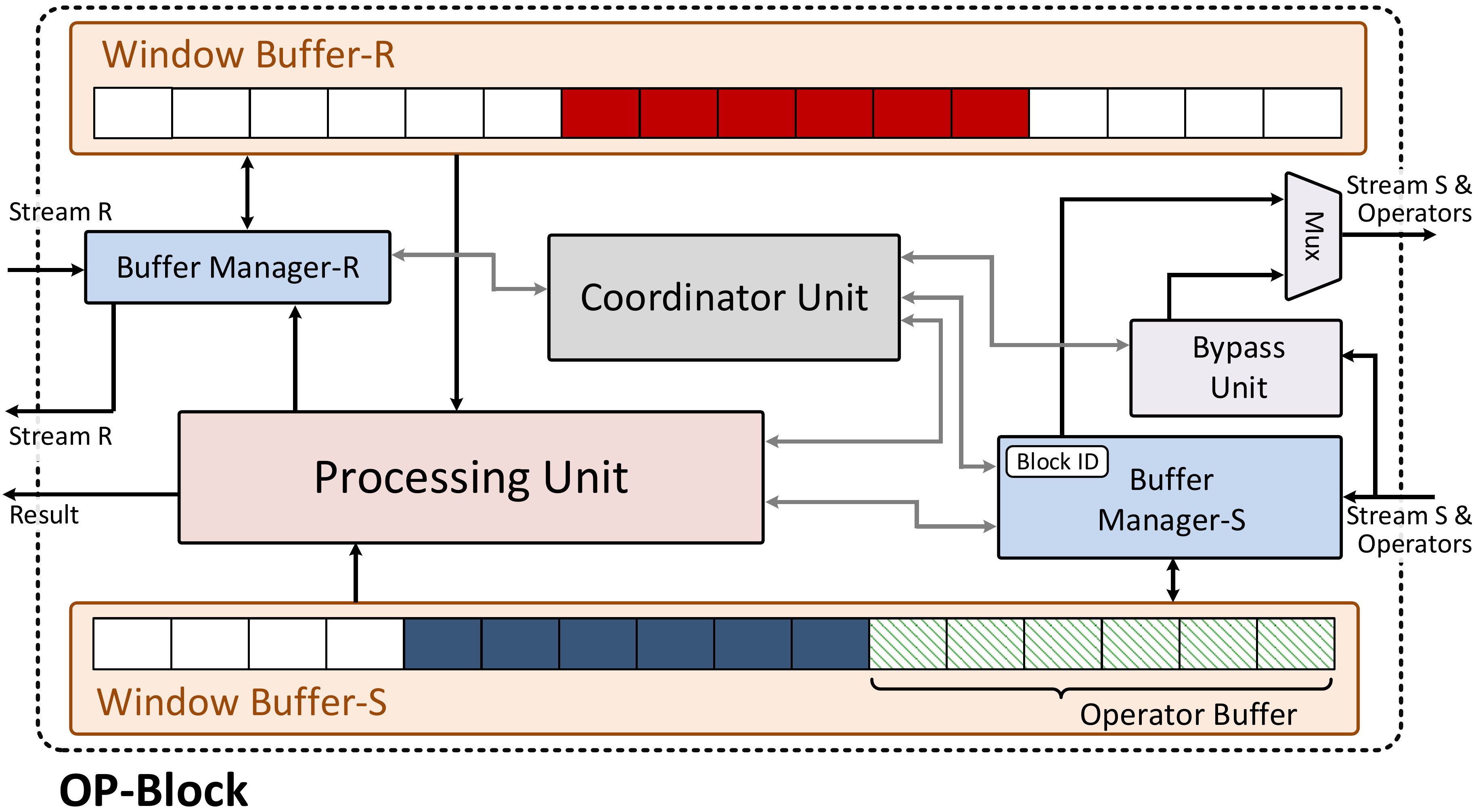}
		\vspace{-5mm}		
		\caption{Original (in \fqp{})}
		\label{fig:original_opblock}	
	\end{subfigure}
	\caption{Internal architecture of an OP-Block}
\end{figure}

Although the \newfqp\ architecture supports and even motivates the use
of general-purpose processing cores (i.e., ARM cores) to handle small
control-intensive tasks within queries, we also have the option of
using custom-programmable processors for the dataflow processing
pattern of the streams. As an example of such processors, we present
our (online programmable block) OP-Block, which is redesigned based on
the unidirectional dataflow processing model and is shown in
Figure~\ref{fig:new_opblock}. The main challenge in the design of
\opb\ is the processing (and parallelization) of the
resource-intensive join operation in stream processing. The current
realization of \opb\ benefits from dedicated memory storage to realize
the sliding-window buffers needed for the join operation, which are
also used as instruction storage.

The original \opb\ and the redesigned (successor) \opb\ are
illustrated in Figures~\ref{fig:original_opblock}
and~\ref{fig:new_opblock}, respectively. In the redesign, data passes
in a single top-down flow, where an \opb\ receives tuples directly
from the \ls. Each \opb\ is placed inside a \pu{} and operates
independently of the other \pus{}. The redesigned architecture fully
utilizes the communication bandwidth provided because all tuples
travel through the same path to the \textit{processing core}, unlike
the bidirectional flow architecture of the original \opb. Therefore,
regardless of the incoming tuple rate for each stream, every tuple has
access to the full bandwidth. The same concept applies when using more
{\opb}s (\pus{}) to parallelize a processing task, as shown in
Figure~\ref{fig:uniflow}.

In the design of \fqp{}, {\opb}s are connected to one another as in
the model shown in Figure~\ref{fig:biflow}, which, in addition to
complicating operations on more than two streams, significantly
degrades performance owing to the underutilized communication
bandwidth. To solve this problem for the original \opb, assume that we
are receiving tuples only from stream $R$; then, all communication
channels for stream $S$ are left unutilized. Even with an equal tuple
rate for both streams, it is impossible to achieve the simultaneous
transmission of both $T_R$ and $T_S$ between neighboring join cores
due to the locks needed to avoid race conditions.

Moreover, the redesigned internal architecture of the \opb{}
(Figure~\ref{fig:new_opblock}) shows a significant reduction in the
number of components and design complexity compared to the original
design (Figure~\ref{fig:original_opblock}). Neighbor-to-neighbor tuple
transfer circuitry for two streams is eliminated from the
\textit{buffer manager-R \& -S} and the \textit{coordination unit}
components. They are reduced and merged to form the \textit{Fetcher}
and \textit{Storage Core} components in the redesigned
architecture. This improvement reduces the number of input/output
ports from five to two, which significantly reduces the hardware
complexity because the number of input/output ports is often an
important indication of the complexity and final cost of a hardware
design. We omit further descriptions of the supported operations and
their instruction sets as they are similar to the ones presented in
~\cite{najafi_fqp_icde}.

\subsection{Complementary Custom Blocks}

We briefly describe the important processing and data handling
components that we designed and implemented (in VHDL) and then
customized for our case study benchmark on \newfqp{}, which is the
TPC-H third query presented in Section~\ref{sec:query_assignment} as a
representative query.

\subsubsection{Multiway Stream Join}

\begin{figure}
	\centering
	\includegraphics[width=1\linewidth]{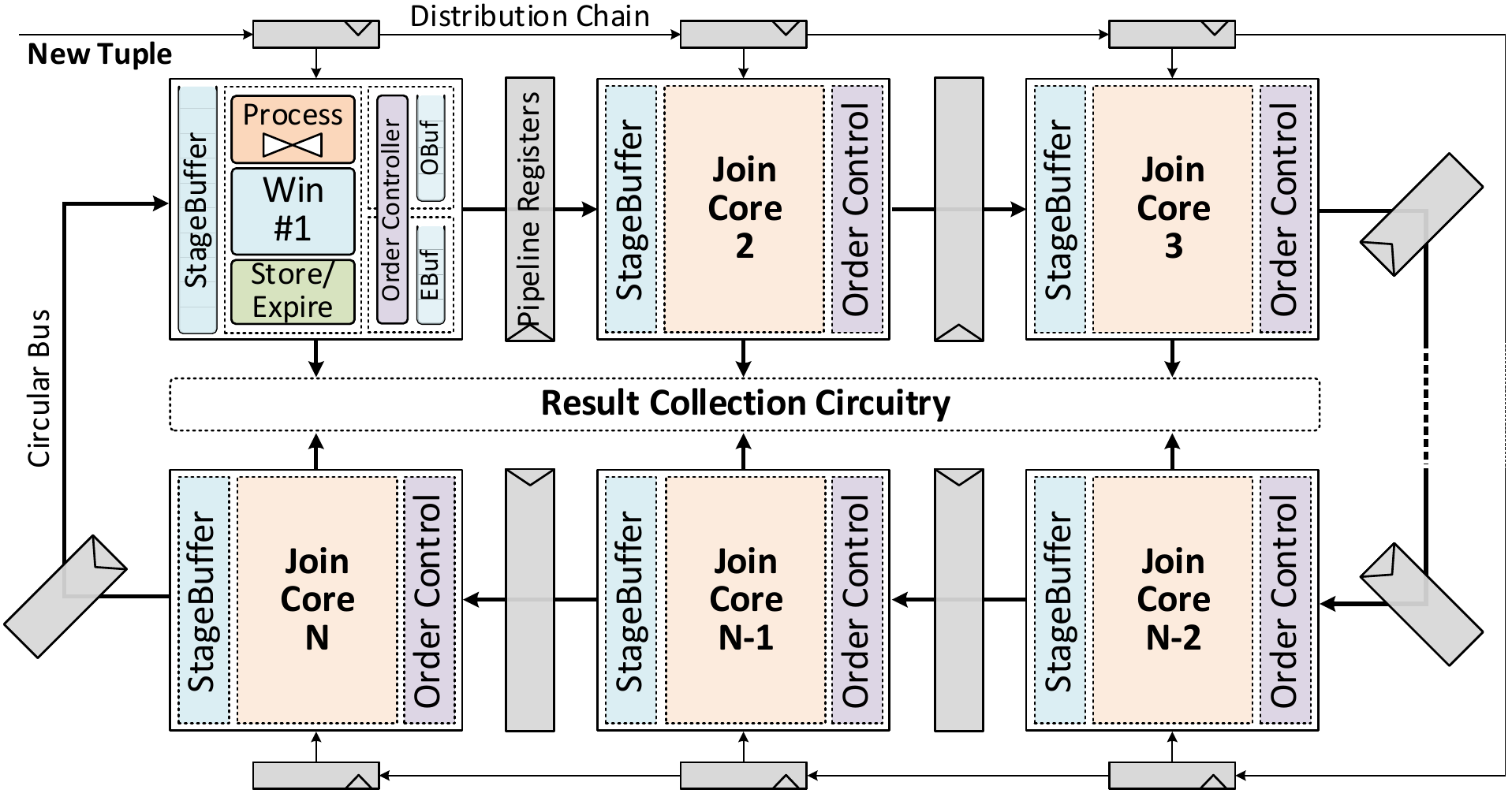}
	\caption{Scalable multiway stream join (\cmj)}
	\label{fig:scale_msjoin}
\end{figure}

Considering the crucial role of joins as among the most
resource-intensive operators in relational databases, it is not
surprising that stream joins have been the focus of considerable
research on data streams~\cite{Kang03_StreamJoin, Gedik09_CellJoin,
  Teubner11_Handshake, Gulisano15_Scalejoin, Lin15_ScalableSJ,
  Wu13_Datapartitioning, Karnagel13_HELLS-join, Oge12_HandshakeFPGA,
  Hagiescu09_FoldingStreamsFPGA}. For example, consider
TPC-H~\cite{Council08_TPC-H}, where 20 queries (of 22) contain a join
operator, and 12 of them use multiway joins, some up to seven
joins. However, the importance of joins is no longer limited to only
the classic relational setting. The emergence of the Internet of
Things (IoT) has introduced a wave of applications that rely on
sensing, gathering, and processing data from an increasingly large
number of connected devices. These applications range from the
scientific and engineering domains to complex pattern matching
methodologies~\cite{Ali05_Nile-PDT, Kumar06_ExpMatching,
  Raghavan07_FireStream}.

In \newfqp, we utilize an instance of our \cmj, presented
in~\cite{Najafi18_CMJ}, placed inside a \pu{} to support multiway
stream joins when needed. \cmj\ benefits from a scalable pipelining
mechanism for processing. The abstract architecture of \cmj\ is shown
in Figure~\ref{fig:scale_msjoin}.

In the design of \cmj{}, each join core (operator) is connected to
only one sliding window with two entries. Each core receives its new
tuples determined based on their origin from its entry that is
connected to a \textit{distribution chain}. Each join core's side
input ports are placed in a circular data path that carries
intermediate results from one join core to the next. The resulting
tuples are emitted after processing a new tuple in exactly $N-1$
cores, where we have $N$ streams. The remaining core is responsible
for storing the new tuple in its sliding window.

The \cmj{} pipeline has the same number of stages (join cores) as
input streams. Each stage is placed between isolating sets of
registers and is responsible for processing a new tuple against a
specific sliding window. When the window in a stage belongs to the
origin of the given tuple, the store and expiration tasks are
performed instead of processing.

In a general \cmj{}, there is a relation between each of the two
streams that require that the circular design has a scalable
architecture; however, in the TPC-H third query, the \textit{lineitem}
and \textit{customer} streams are both related to different attributes
of the \textit{orders'} stream. This property of this query allows for
further customization of our \cmj.

\subsubsection{Hash-based Stream Join}
\begin{figure}
	\centering
	\includegraphics[width=1\linewidth]{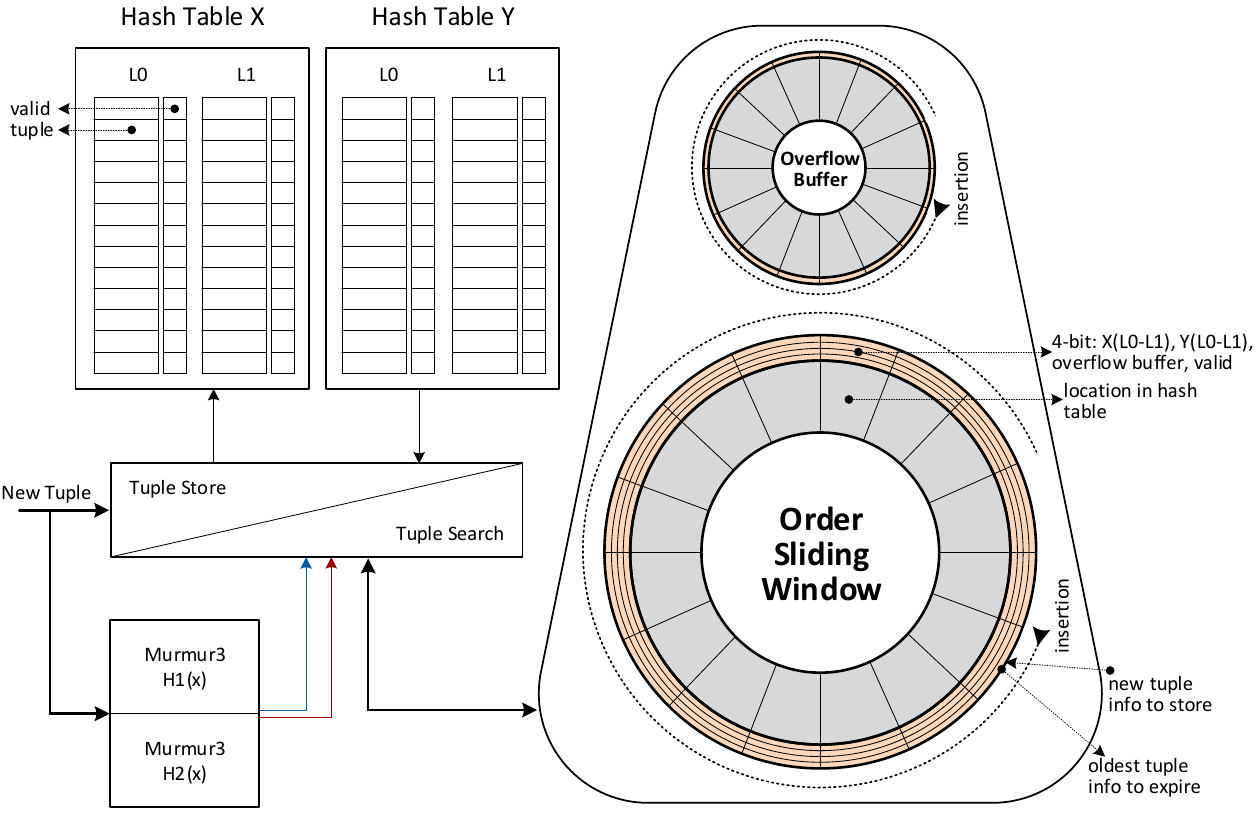}
	\caption{Hash-based architecture for stream join (\hbsj)}
	\label{fig:way-hsj}	
\end{figure}

A nested loop algorithm, although sufficient for the proof of concept
of \newfqp{}, makes it impossible to execute the TPC-H benchmarks in a
reasonably short time because of the sizes of the input data
streams. To accelerate execution, we propose a hardware hash-based
stream join solution that, as the name suggests, benefits from a
customized hashing mechanism to process equi-joins.

Figure~\ref{fig:way-hsj} presents the abstract architecture of our
hash-based stream join unit. It shows only the unit used to process a
stream (e.g., $R$) against an opposing stream (e.g., $S$). For a
complete join, we need to duplicate this architecture for processing
in the other direction. Alternatively, we can use the unit in
Figure~\ref{fig:way-hsj} in each of the join cores in the \cmj\ where,
depending on the number of stages, we can perform anywhere from
two-way up to $N$-way stream joins.

Our hash-based stream join unit, shown in Figure~\ref{fig:way-hsj},
utilizes two Murmur3 hash functions with four storage tables. If the
selected index (row) is full in all four tables, the new tuple is
inserted into an \textit{overflow buffer}. To preserve the order of
arrival of the tuples (necessary in the count-based sliding window),
we use an \textit{ordered sliding window}. The ordered sliding window
keeps the location of each new tuple storage (one of the hash tables
or the overflow buffer) in addition to the hash table index. The
expiration task occurs before inserting a new tuple by removing the
age of the tuple, determined by the oldest tuple in the ordered
sliding window. For a new tuple search, the hash table key is
calculated first, and then all four hash tables are examined in
parallel against the join condition to find their matches. At the end
of the operation, the overflow buffer is searched using a nested loop
join to find matches that are not stored in the hash tables. Although
the overflow buffer is relatively small, we can use a faster algorithm
for the search to accelerate processing as it is still the most
time-consuming operation in our \hbsj.

The architecture of the time-based sliding window does not contain the
\textit{ordered sliding window}, and expiration is performed on new
tuple storage and a new tuple search. Moreover, further adjustments in
the number of hash functions and tables and the use of other
extensions (i.e., the Cuckoo hashing scheme) are beneficial for the
efficiency of hash tables. However, they remain beyond the scope of
\newfqp.

\subsubsection{Aggregation-GroupBy}

\begin{figure}
	\centering
	\begin{subfigure}[t]{0.54\linewidth}
		\centering
		\includegraphics[width=0.85\linewidth]{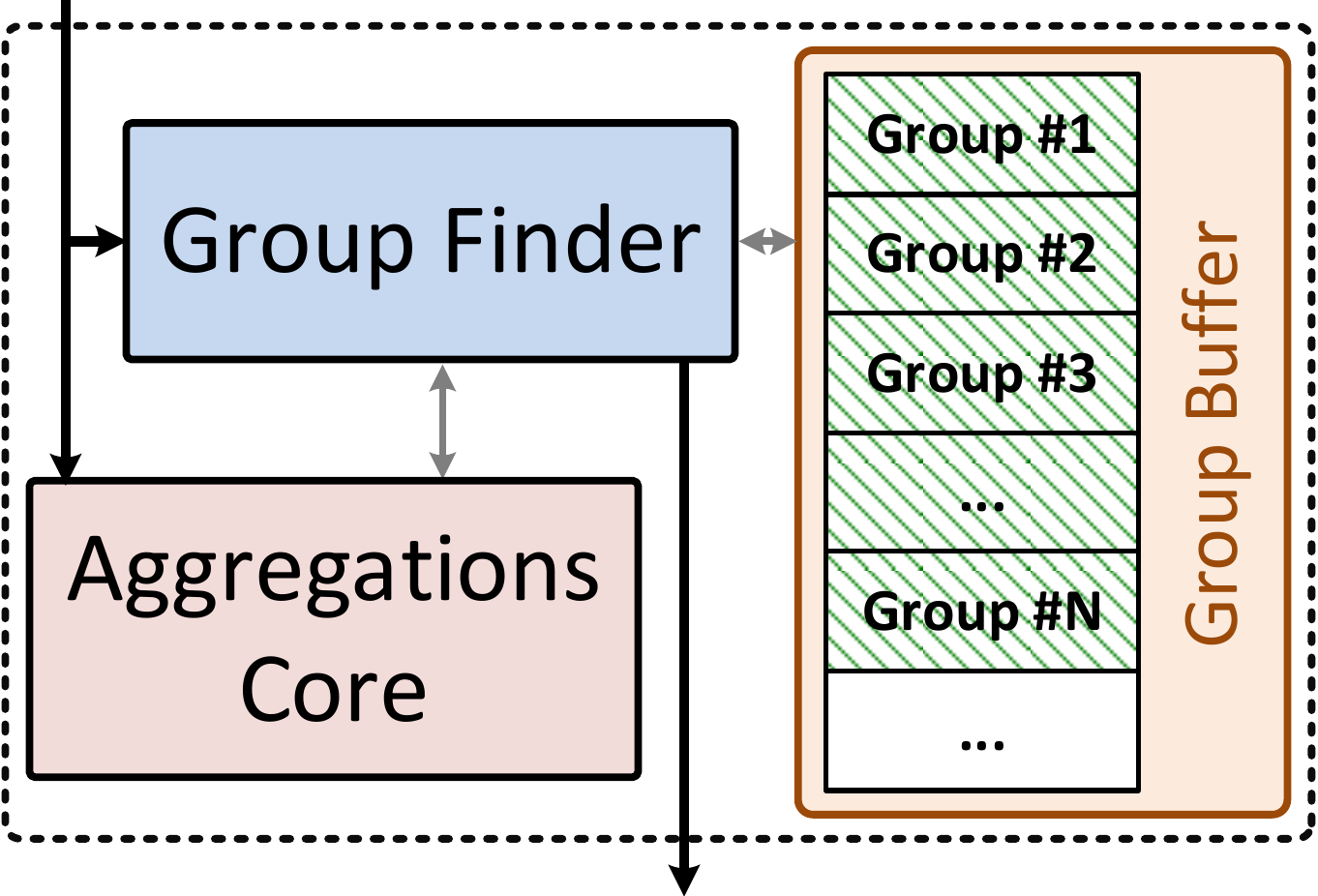}
		\caption{Building blocks}
		\label{fig:agg_groupby}
	\end{subfigure}
	\hfil
	\begin{subfigure}[t]{0.44\linewidth}
		\centering
		\includegraphics[width=1\linewidth]{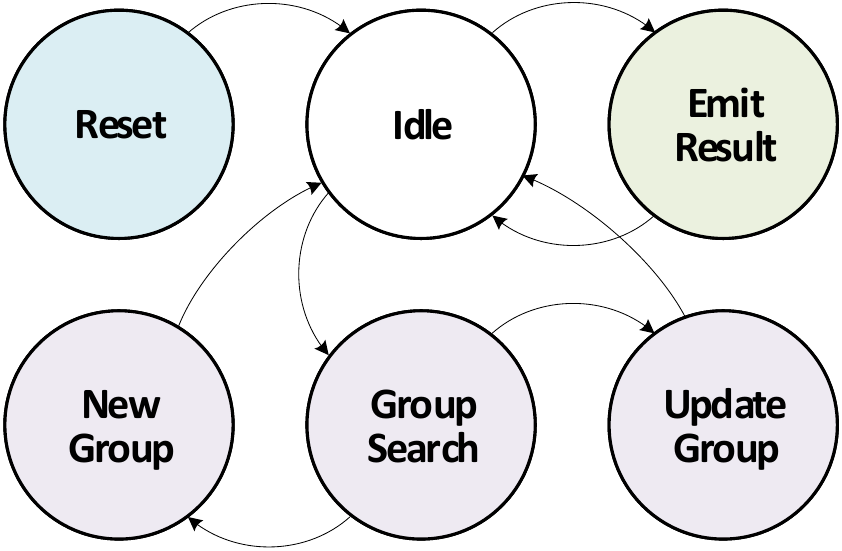}
		\caption{Controller}
		\label{fig:agg_groupby_controller}
	\end{subfigure}
	\caption{\aggregation{}-\groupBy{} unit}
\end{figure}

In queries, it is common to have a grouping operator that is based on
some specified fields while the other fields are aggregated using one
or multiple arithmetic operators. Because the \aggregation{} is
performed on some elements of each group, it is more efficient in the
hardware design to combine these two operators (\aggregation{} and
\groupBy{}) into one unit, as shown in Figure~\ref{fig:agg_groupby}.

Figure~\ref{fig:agg_groupby} shows the building blocks of the
\aggregation{}-\groupBy{} unit, and
Figure~\ref{fig:agg_groupby_controller} presents a simplified
controller state diagram. On inserting a new tuple (or the
intermediate result), the \textit{group finder} unit searches the
\textit{group} buffer to find its relevant group. When a matching
group is found, the aggregation task is performed (in the
{\aggregation{} core}) on that group and the new tuple. The resulting
data are inserted into the group buffer. If no match is found, a new
group is created and filled with the new tuple.

The architecture of the \orderBy{} operator is similar to that of the
\aggregation{}-\groupBy{} unit but with no \aggregation{} core. In the
\orderBy{} operator, the group finder searches for the correct
location of each new group and inserts it. We use bubble sort to
insert each new group into its corresponding sorted location. Other
sorting algorithms can also be implemented in \newfqp{}, which can
improve the performance at the cost of additional hardware complexity.

\subsection{Synchronizer Blocks}

When merging multiple data paths to process a query or set of queries,
a synchronizer component is needed and placed in one of the \pus{}. A
common approach for synchronization is to use tokens generated in a
component, where the correct order of tuples (e.g., from multiple
streams) relative to one another is available. Later in the processing
path, these tokens are used to restore the correct order of tuples
when multiple data paths meet one another.

\subsection{Parallel Processing}

Because \scnoc{} distributes incoming tuples without changing the
order of arrival, we can use parallelization inside or across multiple
topology bricks. The processing throughput for such stateless
operators (i.e., selection) is already high on hardware. However, we
can still parallelize them using a block to reindex a stream into
multiple sub-streams that can be divided by the next \ls\ for
distribution among separate processing blocks.

For more complex state operators such as a join operator, in our
parallelization technique, each processing block performs the
requested task based on its location in the \newfqp\ topology. As an
example, for stream-join parallelization, the architecture of
SplitJoin~\cite{najafi_splitjoin} can be implemented both inside and
across topology bricks. For parallelization inside topology bricks, an
\ls\ replicates and distributes the incoming streams to all processing
units responsible for the join operation, similar to what is shown in
Figure~\ref{fig:uniflow}. Each \pu{} can implement a separate join
core, and depending on its position among the join cores, it stores
one of the multiple received tuples in its respective sub-sliding
window. In contrast, all join cores perform the search task on each
incoming tuple as presented in~\cite{najafi_splitjoin}.

%% file: experimentalresults.tex
\section{Experimental Results}\label{sec:exp}

\begin{figure*}
	\centering
	\includegraphics[width=1.0\linewidth]{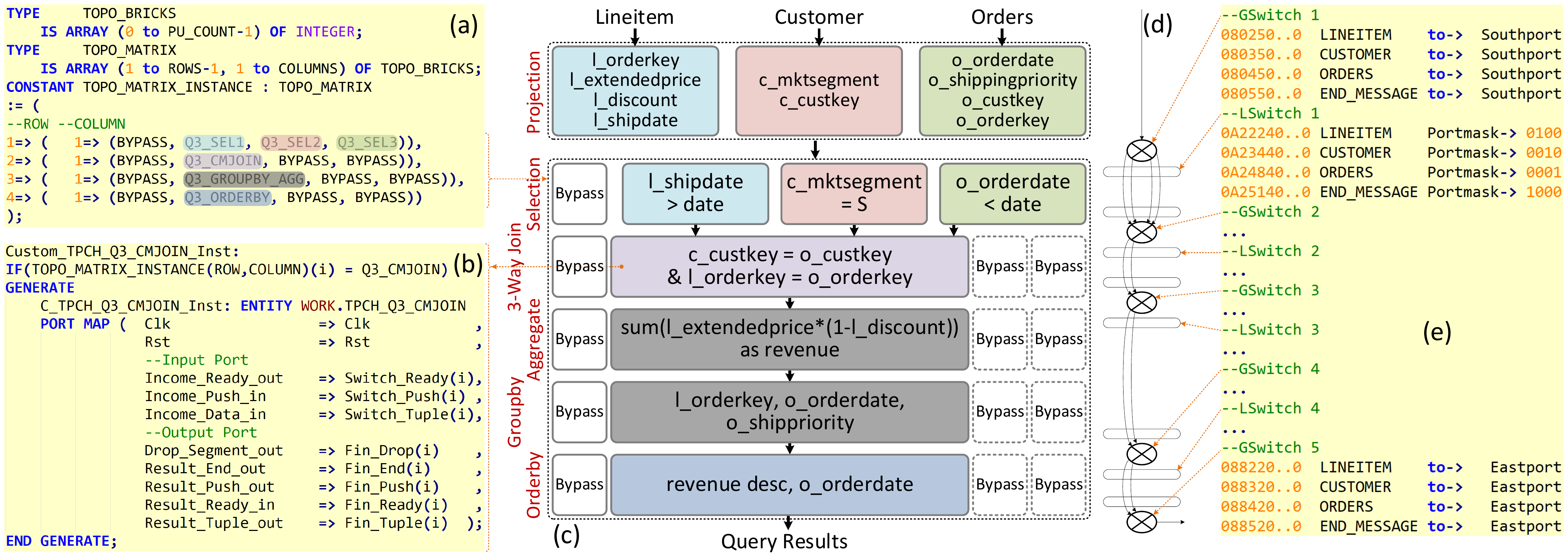}
	\caption{TPC-H third query mapping on \newfqp{}.}
	\vspace{-4mm}
	\label{fig:q3_mapping}	
\end{figure*}

We built a prototype of \newfqp, including its \noc{} and all required
custom blocks, in VHDL. To extract the hardware properties, we
synthesized and implemented our solution on a VCU110 development board
featuring the Virtex UltraScale XCVU190-2FLGC2104E FPGA. We used
Xilinx \textit{Vivado Design Suite} to synthesize and analyze the HDL
design. In the implementation, we used module analysis\footnote{The
  module analysis flow implements a module out-of-context (OOC) of the
  top-level design. The module is implemented in a specific
  part/package combination with a fixed location in the device
  (FPGA). Input/output buffers, global clocks, and other chip-level
  resources are not inserted but can be instantiated within the
  module. The OOC is an important feature that is most useful in large
  hardware systems (designs), where independent synthesis and
  implementation of the modules are necessary to extract the hardware
  properties. Modifying and reimplementing a large-scale system on a
  device takes hours or even days to complete by a synthesis tool. In
  contrast, a significantly shorter time for the synthesis tool is
  needed for each system module owing to its smaller
  size.}~\cite{Vivado_Design_Guide17}, also referred to as
out-of-context implementation, which allows us to analyze a module
independent of the remainder of the design to determine the resource
utilization and extract the timing analysis. We used the
\textit{Questa Advanced Simulator} for functional evaluations to
perform cycle\footnote{Digital hardware operates with pulses of an
  oscillator. The amount of time between pulses is called a clock
  cycle or the inverse of clock frequency. For example, hardware
  operating at a 100-MHz frequency has a clock cycle of 10
  nanoseconds.}-accurate\footnote{A simulation that conforms to the
  cycle-by-cycle behavior of the target hardware.} measurements.

We used the TPC-H DBGen tool~\cite{Council08_TPC-H} to generate the
benchmarking tables and used our parser (written in Python) to
decompose and parse rows of tables into tuples of streams to feed into
\newfqp{}. DBGen allows us to roughly choose the database size
(tables) by a scale factor (SF) parameter. In this work, we used SF=1
(equivalent to a database size of ~1 GB) as the default unless
otherwise stated. In the following, we present a prototype of the
TPC-H third query on the \newfqp{} framework for the evaluations in
this work.

\subsection{TPC-H Third Query Prototype}\label{sec:query_assignment}

To evaluate the functionality and properties of \newfqp{}, we describe in detail the query mapping (to the processing blocks) and programming steps to implement the TPC-H third query (shipping priority). This query retrieves the shipping priority and potential revenue, defined as the \textit{sum of} {\footnotesize \texttt{l\_extendedprice $\times$ (1-l\_discount)}}, of the orders with the largest revenue of those not shipped by a given date. Orders are listed in decreasing order of revenue. If there are more than ten unshipped orders, only ten orders with the largest revenue are listed. Further details are available in the TPC-H benchmark
standard specification manual~\cite{Council08_TPC-H}.

\subsubsection{TPC-H Query Modification}

To execute the third query on \newfqp, we need to change the concept
of the tables to sliding windows and feed each row in a table as a
tuple to our stream processing engine. In queries with state-full
operators, various sliding window sizes and the order of the reception
of new tuples with respect to the origins of the streams affect the
results. This property of stream processing engines needs to be
accounted for in practical use cases, and studying the effects of
these factors is beyond the scope of this work.

\begin{figure}
	\centering
	\includegraphics[width=1.0\linewidth]{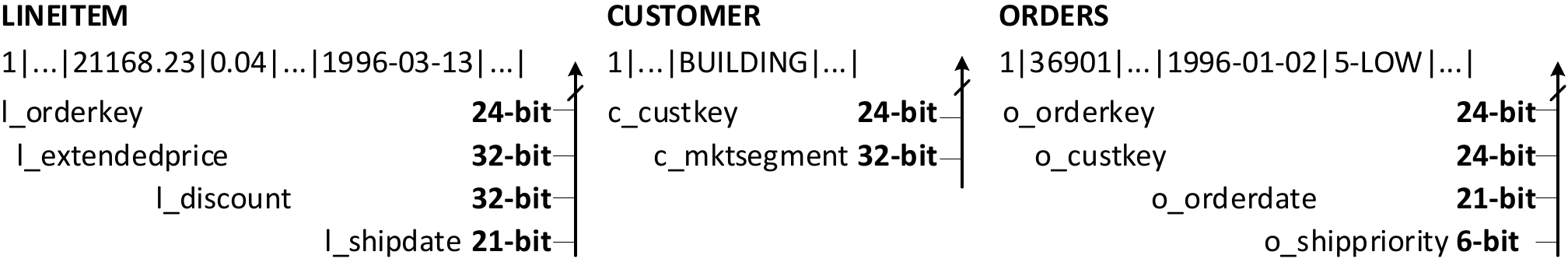}
	\caption{Tuple-to-bit conversion.}
	\label{fig:streams_to_bits}	
\end{figure}

To feed the rows of tables as tuples to \newfqp{}, we allocate a
sufficient number of bits per field in the processing. In
Figure~\ref{fig:streams_to_bits}, we show the mapping of the
attributes of the third query table into different fields, where
several bits specify each attribute field. Subsequently, the total
width of a tuple for each stream is calculated by the sum of the bits
of its fields.

\subsubsection{Query Assignment}

The choice of the topology of \newfqp\ is flexible and can be
determined depending on the system requirements. In the following, we
present an example of the TPC-H third query assignment to a
\newfqp\ instance drawn in detail in Figure~\ref{fig:q3_mapping}. The
assigned components are color-coded to demonstrate the mapping between
the VHDL implementation and the \newfqp\ topology.

The \newfqp\ instance for this example has four topology bricks
arranged in four rows and one column. Note that there is no limit on
the size and form of the chosen topology. In
Figure~\ref{fig:q3_mapping}a, the indexing values defined by
{\footnotesize \texttt{ROW}} and {\footnotesize \texttt{COLUMN}} refer
to each topology brick. For example, the first brick contains a bypass
unit, a selection ({\footnotesize \texttt{Q3\_SEL1}}) on the line-item
stream (built from its equivalent table), a selection ({\footnotesize
  \texttt{Q3\_SEL2}}) on the customer stream, and a selection
({\footnotesize \texttt{Q3\_SEL3}}) on the orders' stream.

A bypass unit is a pass-through operator without any internal
component, and having one of these units in each topology brick is
necessary to transfer instructions to the next row brick. The bypass
unit prevents additional complications in the logic of other
processing units by offloading the unrelated task of bypassing streams
and instructions. Therefore, a topology brick can contain at least two
units, while the upper limit is defined by the application
requirements and fan-out limitations of the chosen hardware.

The implementation of {\footnotesize
  \texttt{TOPO\_MATRIX\-\_INSTANCE}} in VHDL,
Figure~\ref{fig:q3_mapping}a, includes the identifiers of processing
units instantiated subsequently using the VHDL GENERATE construct, as
shown in Figure~\ref{fig:q3_mapping}b. For example, the {\footnotesize
  \texttt{C\_TPCH\-\_Q3\_C\-M\-JOIN\_Inst}} hardware unit is
instantiated and connected to the remaining components because of its
identifier ({\footnotesize \texttt{Q3\_\-CMJOIN}}), in
Figure~\ref{fig:q3_mapping}a. Also, as shown in
Figure~\ref{fig:q3_mapping}a, we have four slots for each row-column
element, which implies that $N=4$ for the \ls{}es used in this
implementation.

Only the necessary fields are fed into the hardware from each tuple to
avoid transmitting/handling unnecessary data. This part implements the
projection operation as a task of the decomposer component (presented
in Figure~\ref{fig:fqpv2completesystem}) in the final system. The
remaining operators are placed in the \newfqp{} instance, as shown in
Figure~\ref{fig:q3_mapping}c, except for the \aggregation{} and
\groupBy{} operators, which were merged together in one unit,
        {\footnotesize \texttt{Q3\_G\-ROUPBY\_AGG}}, in
        Figure~\ref{fig:q3_mapping}a.

To simplify the representation, we show the communication network
(\scnoc) separate from the \newfqp{}'s processing units in
Figure~\ref{fig:q3_mapping}d. The instructions for programming the
\scnoc\ for the TPC-H third query are shown in
Figure~\ref{fig:q3_mapping}e. These instructions are fed into the
\newfqp\ hardware one by one to define the paths of the streams in
each \gs\ and \ls. In addition to the third query input streams, we
define a tuple to indicate the end of the streams, referred to as
{\footnotesize \texttt{END\_MESSAGE}}. The end tuple is fed into the
\newfqp\ hardware after the end of all streams. This tuple is
necessary for the \groupBy{} and \orderBy{} operators to notify them
of the end of the operation, pushing the resulting data from them.

\subsubsection{Multiway Hash-based Stream Joins}

\begin{figure}[t]
	\centering
	\begin{subfigure}[t]{0.8\linewidth}
		\centering
		\includegraphics[width=1\linewidth]{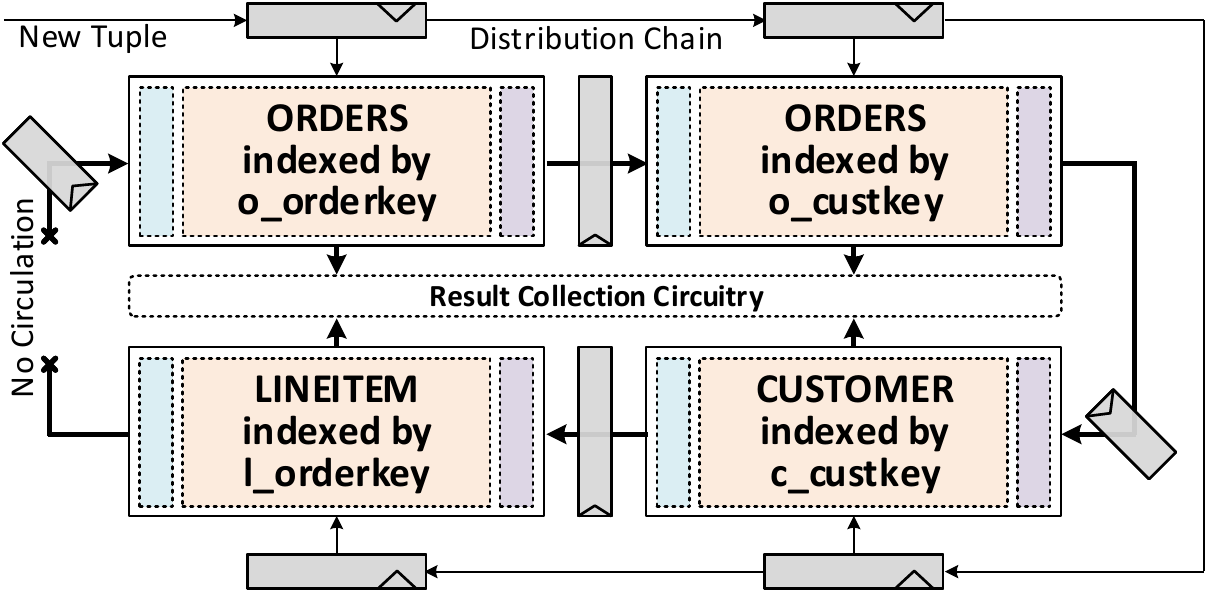}
		\caption{Direct assignment}
		\label{fig:q3_cmjoin1}
	\end{subfigure}
	
	\begin{subfigure}[t]{0.8\linewidth}
		\centering
		\includegraphics[width=1\linewidth]{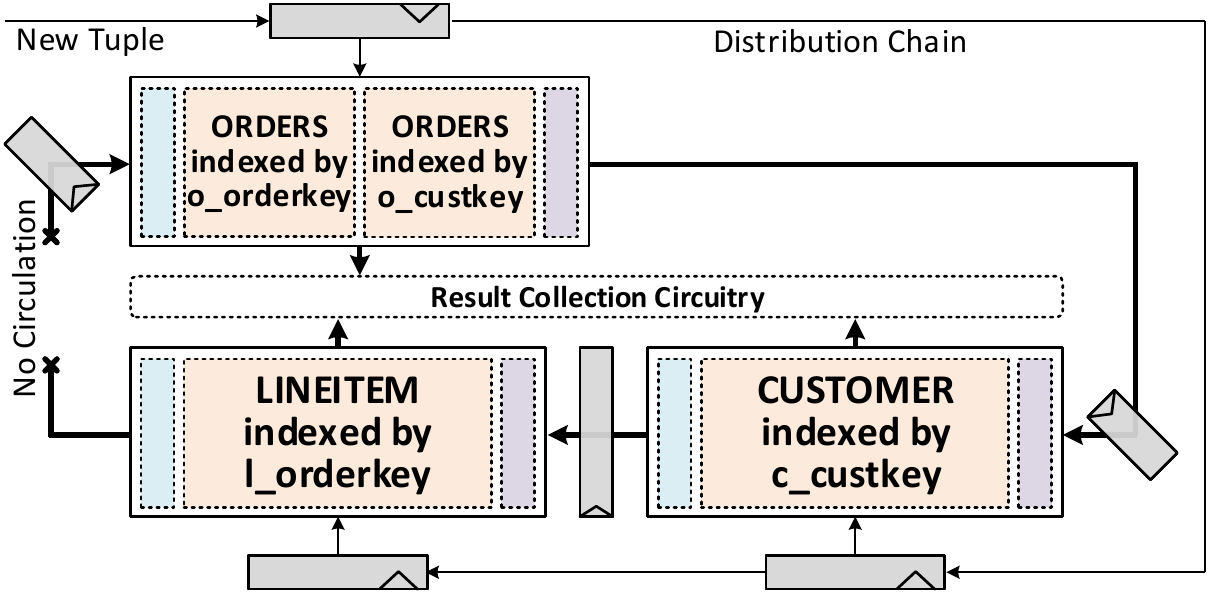}
		\caption{Optimized}
		\label{fig:q3_cmjoin2}
	\end{subfigure}
	\caption{\cmj\ architecture customized for TPC-H 3rd query.}
\end{figure}

We use our hash-based join in a custom three-way stream join for the
third query to demonstrate the potentially high throughput that can be
achieved with \newfqp{}. Using our hash-based approach, we need to
replicate the {\footnotesize \texttt{ORDERS}} stream stage, once
indexed, based on the {\footnotesize \texttt{o\_custkey}} field, and
once based on the {\footnotesize \texttt{o\_orderkey}} field. The
resulting architecture, the customized version of
\cmj\ (Figure~\ref{fig:scale_msjoin}), is shown in
Figure~\ref{fig:q3_cmjoin1}. Note that, due to the simplification of
the multiway stream join operator in the third query, we do not need
the circular path, which is shown by a disconnection mark in this
figure. Because each incoming tuple was processed in, at most, one of
the indexed sliding windows (either {\footnotesize
  \texttt{o\_custkey}} or {\footnotesize \texttt{o\_orderkey}}), we
can further optimize the design by placing both the indexed order
stream stages into one stage to improve the efficiency of the system,
which leads to the architecture shown in Figure~\ref{fig:q3_cmjoin2}.

In Figure~\ref{fig:q3_cmjoin2}, each new tuple is stored in its
corresponding sliding window. For processing the {\footnotesize
  \texttt{ORDERS}} stream, tuples start execution at the
{\footnotesize \texttt{CUSTOMER}} stage, and the resulting tuples are
emitted from the {\footnotesize \texttt{LINE-ITEM}} stage. The
{\footnotesize \texttt{CUSTOMER}} stream tuples start execution at the
{\footnotesize \texttt{ORDERS}} stage, and the intermediate results of
this stage pass through the {\footnotesize \texttt{CUSTOMER}} stage
without processing. The final results are output from the
{\footnotesize \texttt{LINE-ITEM}} stage. Finally, tuples of the
{\footnotesize \texttt{LINE-ITEM}} stream start processing at the
{\footnotesize \texttt{ORDERS}} stage, and the resulting tuples are
emitted from the {\footnotesize \texttt{CUSTOMER}} stage.

\subsection{Throughput Measurements}
\input{ss_throughput}

\begin{figure}
	\centering
	\begin{tikzpicture}[font=\scriptsize]
	\pgfplotstableread{
		380.22813 
		380.51750 
		382.26299 
		382.99502 
		380.80731 
	}\powerfreqtable

	\begin{axis}[
	title style={align=center}, ticks=both, axis x line = bottom, axis y line = left, axis line style={-|}, enlarge y limits={lower, value=0.1}, enlarge y limits={upper, value=0.22}, ylabel style={align=center},
	nodes near coords = \rotatebox{90}{{\pgfmathprintnumber[fixed zerofill, precision=0]{\pgfplotspointmeta}}},	
	nodes near coords align={vertical}, every node near coord/.append style={font=\tiny, yshift=0.5mm}, 
	ylabel={Clock freq/power\\ratio},	xlabel={Clock Period (ns)}, 
	xtick=data, 
	ymin = 380,	
	ymajorgrids, 
	xticklabels={$10$,$9$,$8$,$7$,$6.5$},	
	every axis legend/.append style={nodes={right}, inner sep = 0.1cm},	
	x tick label style={align=center, yshift=-0.1cm}, enlarge x limits=0.12, 
	width=1.0\linewidth, height=3cm,
	]
	\pgfplotsinvokeforeach {0,...,0}{
		\addplot[color1!50!black,fill=color1] table [x expr=\coordindex, y index=#1] {\powerfreqtable};
	}
	\end{axis}
	\end{tikzpicture}
	\vspace{-4mm}
	\caption{Working clock frequency vs. power consumption.}
	\label{fig:power_freq} 
\end{figure}

When working with database benchmarks like TPC-H, most join operators
use equality predicates (i.e., equijoin), where using hashing
techniques is the natural approach for acceleration. However,
following acceleration using our proposed hash-based solution, the
multiway equi-join operator remains the bottleneck for the TPC-H third
query processing. The severity of this bottleneck is determined by the
effectiveness of the hashing technique utilized. We use a window size
of $w:2^{10}$ and an overflow buffer size of $2^{10}$ for these
experiments.

The effect of the size of the hash table ($ht$) used in the optimized
\cmj\ in Figure~\ref{fig:q3_cmjoin2} on the number of cycles needed to
process all data streams is presented in
Figure~\ref{fig:ss_throughput_htsize}. The product of the cycle count
and cycle period (i.e., 6.5 nanoseconds extracted from the
implementations, Figure~\ref{fig:power_freq}) specifies the processing
time. For example, with $ht=2^{9}$, the entire query execution lasts
968.5 milliseconds ($(149\times10^{6})\times(6.5\times10^{-9})$). As
shown in this figure, an increase in the size of the hash tables, from
left to right, improves processing throughput due to a reduction in
the number of tuple insertion collisions in the hash tables. This
significantly reduces the number of tuples in the overflow buffer
(Figure~\ref{fig:way-hsj}), which uses a slow nested loop search.

\input{ss_tput_dbsize}

The effect of the size of the data stream (equivalent to the size of
the database generated by the DBGen tool) on processing times for the
TPC-H third query is illustrated in
Figure~\ref{fig:ss_throughput_dbsize}. We observe a linear increase in
processing time with the size of the data stream. With a clock period
of 10 nanoseconds (clock frequency of 100 MHz), data streams of sizes
one, two, four, six, eight, or 10 GB were processed in 300
($30\times(10\times10^{-9})\times10^{6}$), 610, 1220, 1850, 2670, and
3520 milliseconds, respectively.

\input{ob_tuples_time}

The effectiveness of our hashing mechanism was determined by the
number of tuples stored in the overflow buffer, as shown in
Figure~\ref{fig:ovbuf_time}. A small number means fewer tuple
insertion collisions and, therefore, more effective hashing, and vice
versa. We show four sub-figures here, and each presents the effect on
a single indexed attribute. This effectiveness was also influenced by
the pattern of the received values and the size of each stream. In
this figure, we observe a slower growth in the number of tuples in the
overflow buffer for the order and customer streams compared with the
line-item stream; this result was obtained owing to the large
difference in the sizes of these streams. In other words, we fed
\newfqp{} with more tuples from the line-item stream compared with the
other two streams.

We observed a similar increase in the use of the overflow buffer as we
reduced the size of the hash table for each stream. An interesting
observation was the slow growth in the use of the overflow buffer for
the {\footnotesize \texttt{o\_cust}} attribute compared with the
{\footnotesize \texttt{o\_orderkey}} attribute (both from the \ord{}
stream) due to the reception of more diverse values for the
{\footnotesize \texttt{o\_cust}} attribute with regard to the order of
arrival of the tuples.

\input{throughput_time}

Figure~\ref{fig:throughput_time} presents variations in the input
throughput with the number of processing cycles for different hash
table sizes ranging from $2^{5}$ to $2^{11}$. By reducing the
effectiveness of the hashing mechanism, more tuples are stored in the
overflow buffer. This translates into a more sequential comparison of
tuples, which significantly reduces processing throughput. As a side
effect, we can also clearly see the warm-up phase in the experiments
with smaller hash tables, which shows longer processing times as a
function of an increase in the number of tuples in the overflow
buffer.

\textbf{Evaluation of Simple Queries.}

The processing of queries that do not have resource-intensive
operators (e.g., join) is considerably faster, especially in hardware.
The TPC-H first query (pricing summary report) is a good example of
such queries. This query reports the volume of business that is
billed, shipped, and returned. In this query, following the
projection, line-item tuples are fed into a selection operator, and
the intermediate results are sent to an \aggregation{} operator. In
the end, the outcome of the \aggregation{} operator is processed by a
\groupBy{} operator and then an \orderBy{} operator. With only a
single processing path, \newfqp{} processed a database of size 1 GB in
less than 22.4 million cycles, which translates into 224 milliseconds
at a clock frequency of 100 MHz. Moreover, in such queries, it is
possible to simply replicate the processing path multiple times to
obtain a linear speedup.

\subsection{Power Consumption Evaluations}

\input{q3_power}

Dynamic power consumption measurements, including the contributions of
different types of FPGA resources, are presented in
Figure~\ref{fig:q3_power}. In these measurements, \textit{clocks}
shows the power consumed by the clock network, and \textit{signals}
shows the power consumed by the interconnections between
components. \textit{Logic} refers to the processing and routing
components realized by FPGA configurable logic blocks
(CLBs). \textit{BRAM} shows the power consumed by memory elements in
the design, and \textit{DSP} presents the power consumed by the
built-in (faster and more efficient) digital signal processing units,
such as float multipliers.

In our case study, the processing throughput improved linearly in
relation to the frequency of the operating clock. The ratios of
processing throughput to power consumption for various working clock
frequencies are presented in Figure~\ref{fig:power_freq}. Although the
difference in these ratios is better shown in ASIC solutions, owing to
their higher clock frequencies and fixed implementation, we still
observe a clear peak at a clock period of 7 nanoseconds ($\sim$142
MHz). In these implementations, we kept the other realization
conditions intact to reduce the chance of changes in the
implementation by using the synthesis tool due to its internal
optimizations. However, opting for a higher clock frequency (e.g.,
clock periods smaller than 6.5 nanoseconds) forced the synthesis tool
to use more intrusive optimization techniques (i.e., replication of
processing logic) to achieve the targeted clock frequency. In
Figure~\ref{fig:q3_power}, we observe a similar consumption for each
category of components across multiple realizations, each with a
different clock frequency. This implies that the synthesis tool
generated similar implementations for various clock frequencies.

\subsection{Evaluating the Implementation}

\input{q3_util}

A detailed summary of the hierarchical resource utilization for the third TPC-H query is presented in Table~\ref{tab:tpchq3_table}. The third TPC-H query is a particularly resource-intensive query due to its multiway stream join operator. However, the implementation of a state-of-the-art FPGA consumes only a fraction of the available resources. This demonstrates the applicability of \newfqp{} to deploy multiple queries with even more complex processing components.

In Table~\ref{tab:tpchq3_table}, we observe low use of block RAM tiles (six of 3,780) for hash tables and their corresponding components in the custom multiway stream join (Figure~\ref{fig:q3_cmjoin2}). In this implementation, we use small sliding windows because the memory resources provided in an FPGA are valuable and limited. When working with large sliding windows, a decision concerning the use of one or multiple external memory chips (directly connected to the FPGA) needs to be made. This decision frees up valuable FPGA memory resources for logic and crucial buffering.

The TPC-H third query used five {\gsl}s (Figure~\ref{fig:q3_mapping}), and they consumed a negligible amount of FPGA resources. Similarly, the {\footnotesize \textsf{{\ls}s}} consumed minimal resources from our FPGA, as observed in the resource utilization report of \ls{} in the second topology brick (\brick{2}). Therefore, \scnoc{} complied with its objectives of serving as a simple and undemanding communication layer.

\brick{1} contained the selection operators and consumed a negligible amount of resources, as expected of stateless operators, whereas \brick{2} consumed the most resources due to its multiway stream join operator. \brick{3} contained the \aggregation{}-\groupBy{} operator, and \brick{4} contained the \orderBy{} operator; however, they also consumed a negligibly small amount of resources compared with the second topology brick.

In Table~\ref{tab:tpchq3_table}, we observe the resource consumption in the implementation of the pipeline stages for our custom multiway stream joins. {PipeStage2} used double the resources for the other two stages because it contained two separate instances of our \hbsj, although sharing some components resulted in a slight reduction in terms of resource consumption.

%% file: ss_throughput.tex
\begin{figure}
	\centering
	\begin{tikzpicture}[font=\scriptsize]
	\pgfplotstableread{
		1723.308362
		1502.788926
		1123.365921
		581.268706
		149.244137
		37.061318
		30.468251
	}\datatable
	
	\begin{axis}[ybar, ybar=0pt, 
	xlabel near ticks,
	ylabel near ticks,
	bar width = 0.22cm, title style={align=center}, ticks=both, axis x line = bottom, axis y line = left, axis line style={-|}, enlarge y limits={lower, value=0.1}, enlarge y limits={upper, value=0.22}, ylabel style={align=center},
	nodes near coords = \rotatebox{90}{{\pgfmathprintnumber[fixed zerofill, precision=0]{\pgfplotspointmeta}}},	nodes near coords align={vertical},	every node near coord/.append style={font=\tiny, yshift=0.5mm}, 
	ylabel={Processing Cycles\\($\times$million)},	xlabel=Hash Table Size ($2^x$), xtick=data, ymin = 0,	ymajorgrids, xticklabels={$5$,$6$,$7$,$8$,$9$,$10$,$11$},	legend style={at={(0.5, 1.0)}, anchor=north, legend columns=4},
	every axis legend/.append style={nodes={right}, inner sep = 0.1cm},	x tick label style={align=center, yshift=-0.1cm}, enlarge x limits=0.12, width=1.0\linewidth, height=3cm,
	]
	\pgfplotsinvokeforeach {0,...,0}{
		\addplot[color1!50!black,fill=color1] table [x expr=\coordindex, y index=#1] {\datatable};
	}
	\end{axis}
	\end{tikzpicture}
	\vspace{-6mm}
	\caption{Input throughput vs. hash table size ($w:2^{10}$).}
	\label{fig:ss_throughput_htsize} 
\end{figure}

%% file: ss_tput_dbsize.tex
\begin{figure}
	\centering
	\begin{tikzpicture}[font=\scriptsize]
	\pgfplotstableread{
		30.468251
		60.570657
		121.759095
		185.278114
		266.953056
		351.542534
	}\datatable
	
	\begin{axis}[ybar, ybar=0pt, bar width = 0.22cm, title style={align=center}, ticks=both, axis x line = bottom, axis y line = left, axis line style={-|}, enlarge y limits={lower, value=0.1}, enlarge y limits={upper, value=0.22}, ylabel style={align=center},
	nodes near coords = \rotatebox{90}{{\pgfmathprintnumber[fixed zerofill, precision=0]{\pgfplotspointmeta}}},	nodes near coords align={vertical},	every node near coord/.append style={font=\tiny, yshift=0.5mm}, 
	ylabel={Processing Cycles\\($\times$million)},	xlabel=Database Size ($GB$), xtick=data, ymin = 0,	ymajorgrids, xticklabels={$1$,$2$,$4$,$6$,$8$,$10$},	legend style={at={(0.5, 1.0)}, anchor=north, legend columns=4},
	every axis legend/.append style={nodes={right}, inner sep = 0.1cm},	x tick label style={align=center, yshift=-0.1cm}, enlarge x limits=0.12, width=1.0\linewidth, height=3cm,
	]
	\pgfplotsinvokeforeach {0,...,0}{
		\addplot[color1!50!black,fill=color1] table [x expr=\coordindex, y index=#1] {\datatable};
	}
	\end{axis}
	\end{tikzpicture}
	\vspace{-4mm}
	\caption{Input throughput vs. database size ($w:2^{10}, ht:2^{11}$).}
	\label{fig:ss_throughput_dbsize} 
\end{figure}
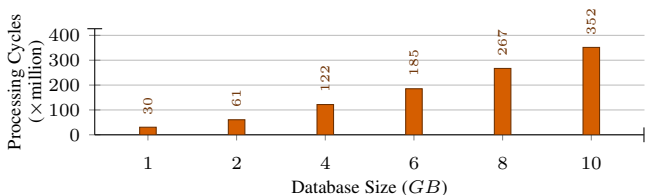

%% file: ob_tuples_time.tex
\begin{figure}
	\begin{tikzpicture}[font=\scriptsize]
	\begin{groupplot}[
	group style={
		group name=my plots,
		group size=1 by 4,
		xlabels at=edge bottom,
		xticklabels at=edge bottom,
		vertical sep=2pt
	},
	width=1.0\linewidth,	height=2.8cm,	xmajorgrids, xlabel=Processing Cycles ($\times 10^{5}$), xtick distance=30,	enlarge x limits=0.05,	scaled y ticks = false,	legend style={at={(0.45, 1.45)}, anchor=north, legend columns=8}, xmax = 280, ytick distance=200, ymax=990, tickpos=left, ytick align=outside, xtick align=outside, legend style={font=\scriptsize,text height=.5em},
	]
	\nextgroupplot [ylabel=Number of tuples in the overflow buffer,  y label style={at={(-0.09,-1.0)}},title=($l\_orderkey$),title style={font=\footnotesize,xshift=0, yshift=-4em}]	
	\addlegendimage{empty legend}
	\addlegendentry{$ht:$}; 
	\addplot[color1,very thick, no marks, select coords between index={0}{120}] plot table[x=sample_avg,y=s1_buf_avg]{AV10_W0_U_M4_D2_HB_S3_W2_10_R10_6_RP10_5_PB16384_HT32_ST1024_OVERFLOWBUF.dat};
	\addlegendentry{$2^{5}$}; 	
	\addplot[color2,very thick, no marks, select coords between index={0}{120}] plot table[x=sample_avg,y=s1_buf_avg]{AV10_W0_U_M4_D2_HB_S3_W2_10_R10_6_RP10_5_PB16384_HT64_ST1024_OVERFLOWBUF.dat};		
	\addlegendentry{$2^{6}$};	
	\addplot[color3,very thick, no marks, select coords between index={0}{120}] plot table[x=sample_avg,y=s1_buf_avg]{AV10_W0_U_M4_D2_HB_S3_W2_10_R10_6_RP10_5_PB16384_HT128_ST1024_OVERFLOWBUF.dat};		
	\addlegendentry{$2^{7}$};	
	\addplot[color4,very thick, no marks, select coords between index={0}{120}] plot table[x=sample_avg,y=s1_buf_avg]{AV10_W0_U_M4_D2_HB_S3_W2_10_R10_6_RP10_5_PB16384_HT256_ST1024_OVERFLOWBUF.dat};		
	\addlegendentry{$2^{8}$};	
	\addplot[color5,very thick, no marks, select coords between index={0}{120}] plot table[x=sample_avg,y=s1_buf_avg]{AV10_W0_U_M4_D2_HB_S3_W2_10_R10_6_RP10_5_PB16384_HT512_ST1024_OVERFLOWBUF.dat};		
	\addlegendentry{$2^{9}$};	
	\addplot[color6,very thick, no marks, select coords between index={0}{120}] plot table[x=sample_avg,y=s1_buf_avg]{AV10_W0_U_M4_D2_HB_S3_W2_10_R10_6_RP10_5_PB16384_HT1024_ST1024_OVERFLOWBUF.dat};	
	\addlegendentry{$2^{10}$};	
	\addplot[color7,very thick, no marks, select coords between index={0}{120}] plot table[x=sample_avg,y=s1_buf_avg]{AV10_W0_U_M4_D2_HB_S3_W2_10_R10_6_RP10_5_PB16384_HT2048_ST1024_OVERFLOWBUF.dat};	
	\addlegendentry{$2^{11}$};	
	
	\nextgroupplot[title=($o\_orderkey$),title style={font=\footnotesize,xshift=0, yshift=-4em}]
	\addplot[color1,very thick, no marks, select coords between index={0}{120}] plot table[x=sample_avg,y=s2_buf_avg]{AV10_W0_U_M4_D2_HB_S3_W2_10_R10_6_RP10_5_PB16384_HT32_ST1024_OVERFLOWBUF.dat};	
	\addplot[color2,very thick, no marks, select coords between index={0}{120}] plot table[x=sample_avg,y=s2_buf_avg]{AV10_W0_U_M4_D2_HB_S3_W2_10_R10_6_RP10_5_PB16384_HT64_ST1024_OVERFLOWBUF.dat};		
	\addplot[color3,very thick, no marks, select coords between index={0}{120}] plot table[x=sample_avg,y=s2_buf_avg]{AV10_W0_U_M4_D2_HB_S3_W2_10_R10_6_RP10_5_PB16384_HT128_ST1024_OVERFLOWBUF.dat};		
	\addplot[color4,very thick, no marks, select coords between index={0}{120}] plot table[x=sample_avg,y=s2_buf_avg]{AV10_W0_U_M4_D2_HB_S3_W2_10_R10_6_RP10_5_PB16384_HT256_ST1024_OVERFLOWBUF.dat};		
	\addplot[color5,very thick, no marks, select coords between index={0}{120}] plot table[x=sample_avg,y=s2_buf_avg]{AV10_W0_U_M4_D2_HB_S3_W2_10_R10_6_RP10_5_PB16384_HT512_ST1024_OVERFLOWBUF.dat};		
	\addplot[color6,very thick, no marks, select coords between index={0}{120}] plot table[x=sample_avg,y=s2_buf_avg]{AV10_W0_U_M4_D2_HB_S3_W2_10_R10_6_RP10_5_PB16384_HT1024_ST1024_OVERFLOWBUF.dat};	
	\addplot[color7,very thick, no marks, select coords between index={0}{120}] plot table[x=sample_avg,y=s2_buf_avg]{AV10_W0_U_M4_D2_HB_S3_W2_10_R10_6_RP10_5_PB16384_HT2048_ST1024_OVERFLOWBUF.dat};		
	
	\nextgroupplot[title=($c\_cust$),title style={font=\footnotesize,xshift=0, yshift=-4em}]
	\addplot[color1,very thick, no marks, select coords between index={0}{120}] plot table[x=sample_avg,y=s3_buf_avg]{AV10_W0_U_M4_D2_HB_S3_W2_10_R10_6_RP10_5_PB16384_HT32_ST1024_OVERFLOWBUF.dat};	
	\addplot[color2,very thick, no marks, select coords between index={0}{120}] plot table[x=sample_avg,y=s3_buf_avg]{AV10_W0_U_M4_D2_HB_S3_W2_10_R10_6_RP10_5_PB16384_HT64_ST1024_OVERFLOWBUF.dat};		
	\addplot[color3,very thick, no marks, select coords between index={0}{120}] plot table[x=sample_avg,y=s3_buf_avg]{AV10_W0_U_M4_D2_HB_S3_W2_10_R10_6_RP10_5_PB16384_HT128_ST1024_OVERFLOWBUF.dat};		
	\addplot[color4,very thick, no marks, select coords between index={0}{120}] plot table[x=sample_avg,y=s3_buf_avg]{AV10_W0_U_M4_D2_HB_S3_W2_10_R10_6_RP10_5_PB16384_HT256_ST1024_OVERFLOWBUF.dat};		
	\addplot[color5,very thick, no marks, select coords between index={0}{120}] plot table[x=sample_avg,y=s3_buf_avg]{AV10_W0_U_M4_D2_HB_S3_W2_10_R10_6_RP10_5_PB16384_HT512_ST1024_OVERFLOWBUF.dat};		
	\addplot[color6,very thick, no marks, select coords between index={0}{120}] plot table[x=sample_avg,y=s3_buf_avg]{AV10_W0_U_M4_D2_HB_S3_W2_10_R10_6_RP10_5_PB16384_HT1024_ST1024_OVERFLOWBUF.dat};	
	\addplot[color7,very thick, no marks, select coords between index={0}{120}] plot table[x=sample_avg,y=s3_buf_avg]{AV10_W0_U_M4_D2_HB_S3_W2_10_R10_6_RP10_5_PB16384_HT2048_ST1024_OVERFLOWBUF.dat};		
	
	\nextgroupplot[title=($o\_cust$),title style={font=\footnotesize,xshift=0, yshift=-4em}]
	\addplot[color1,very thick, no marks, select coords between index={0}{120}] plot table[x=sample_avg,y=s4_buf_avg]{AV10_W0_U_M4_D2_HB_S3_W2_10_R10_6_RP10_5_PB16384_HT32_ST1024_OVERFLOWBUF.dat};	
	\addplot[color2,very thick, no marks, select coords between index={0}{120}] plot table[x=sample_avg,y=s4_buf_avg]{AV10_W0_U_M4_D2_HB_S3_W2_10_R10_6_RP10_5_PB16384_HT64_ST1024_OVERFLOWBUF.dat};		
	\addplot[color3,very thick, no marks, select coords between index={0}{120}] plot table[x=sample_avg,y=s4_buf_avg]{AV10_W0_U_M4_D2_HB_S3_W2_10_R10_6_RP10_5_PB16384_HT128_ST1024_OVERFLOWBUF.dat};		
	\addplot[color4,very thick, no marks, select coords between index={0}{120}] plot table[x=sample_avg,y=s4_buf_avg]{AV10_W0_U_M4_D2_HB_S3_W2_10_R10_6_RP10_5_PB16384_HT256_ST1024_OVERFLOWBUF.dat};		
	\addplot[color5,very thick, no marks, select coords between index={0}{120}] plot table[x=sample_avg,y=s4_buf_avg]{AV10_W0_U_M4_D2_HB_S3_W2_10_R10_6_RP10_5_PB16384_HT512_ST1024_OVERFLOWBUF.dat};		
	\addplot[color6,very thick, no marks, select coords between index={0}{120}] plot table[x=sample_avg,y=s4_buf_avg]{AV10_W0_U_M4_D2_HB_S3_W2_10_R10_6_RP10_5_PB16384_HT1024_ST1024_OVERFLOWBUF.dat};	
	\addplot[color7,very thick, no marks, select coords between index={0}{120}] plot table[x=sample_avg,y=s4_buf_avg]{AV10_W0_U_M4_D2_HB_S3_W2_10_R10_6_RP10_5_PB16384_HT2048_ST1024_OVERFLOWBUF.dat};		
	
	\end{groupplot}
	
	\end{tikzpicture}
        \vspace{-4mm}			
	\caption{Number of tuples in the overflow buffer ($w:2^{10}$).} 
	\label{fig:ovbuf_time}
\end{figure}
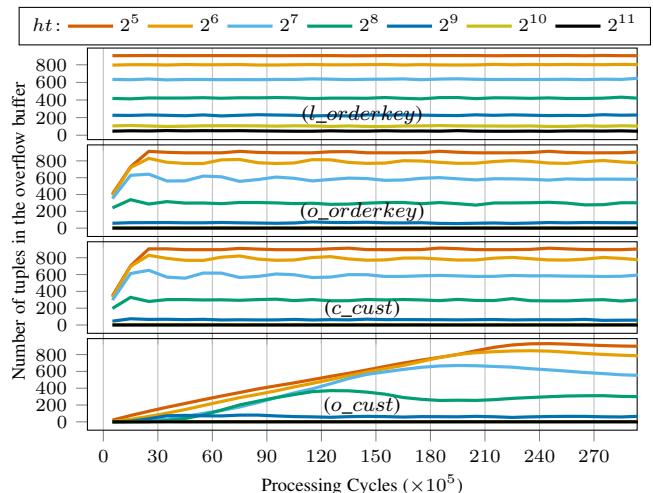

%% file: throughput_time.tex
\begin{figure}
	\begin{tikzpicture}[font=\scriptsize]
	\begin{groupplot}[
	group style={
		group name=my plots,
		group size=1 by 7,
		xlabels at=edge bottom,
		xticklabels at=edge bottom,
		vertical sep=1pt
	},
	width=1.0\linewidth,	height=2.6cm,	xmajorgrids, xlabel=Processing Cycles ($\times 10^{5}$), xtick distance=30,	enlarge x limits=0.05,	scaled y ticks = false,	legend style={at={(0.5, 1.5)}, anchor=north, legend columns=3}, xmax = 280, tickpos=left, ytick align=outside, xtick align=outside, legend style={font=\scriptsize,text height=.5em}]
	
	\nextgroupplot[ylabel=Throughput ($tuples/100000\ cycles$),  y label style={at={(-0.11,-3.0)}},title=($ht$:$2^{11}$),title style={font=\footnotesize,xshift=0, yshift=-3em}]
	\addplot+[red,thick, dotted, no marks, select coords between index={5}{120}]	plot table[x=sample_avg,y=input_max]{AV10_W0_U_M4_D2_HB_S3_W2_10_R10_6_RP10_5_PB16384_HT2048_ST1024_THROUGHPUT.dat};
	\addlegendentry{$Max$}; 	
	\addplot+[black,very thick, no marks, select coords between index={5}{120}] plot table[x=sample_avg,y=input_avg]{AV10_W0_U_M4_D2_HB_S3_W2_10_R10_6_RP10_5_PB16384_HT2048_ST1024_THROUGHPUT.dat};	
	\addlegendentry{$Avg$}; 
	\addplot+[blue,thick, dotted, no marks, select coords between index={5}{120}]	plot table[x=sample_avg,y=input_min]{AV10_W0_U_M4_D2_HB_S3_W2_10_R10_6_RP10_5_PB16384_HT2048_ST1024_THROUGHPUT.dat};	
	\addlegendentry{$Min$}; 
	
	\nextgroupplot [title=($ht$:$2^{10}$),title style={font=\footnotesize,xshift=0, yshift=-3em}]
	\addplot+[red,thick, dotted, no marks, select coords between index={5}{120}]	plot table[x=sample_avg,y=input_max]{AV10_W0_U_M4_D2_HB_S3_W2_10_R10_6_RP10_5_PB16384_HT1024_ST1024_THROUGHPUT.dat};
	\addplot+[black,very thick, no marks, select coords between index={5}{120}] plot table[x=sample_avg,y=input_avg]{AV10_W0_U_M4_D2_HB_S3_W2_10_R10_6_RP10_5_PB16384_HT1024_ST1024_THROUGHPUT.dat};	
	\addplot+[blue,thick, dotted, no marks, select coords between index={5}{120}]	plot table[x=sample_avg,y=input_min]{AV10_W0_U_M4_D2_HB_S3_W2_10_R10_6_RP10_5_PB16384_HT1024_ST1024_THROUGHPUT.dat};	

	\nextgroupplot [title=($ht$:$2^{9}$),title style={font=\footnotesize,xshift=0, yshift=-3em}]
	\addplot+[red,thick, dotted, no marks, select coords between index={5}{120}]	plot table[x=sample_avg,y=input_max]{AV10_W0_U_M4_D2_HB_S3_W2_10_R10_6_RP10_5_PB16384_HT512_ST1024_THROUGHPUT.dat};
	\addplot+[black,very thick, no marks, select coords between index={5}{120}] plot table[x=sample_avg,y=input_avg]{AV10_W0_U_M4_D2_HB_S3_W2_10_R10_6_RP10_5_PB16384_HT512_ST1024_THROUGHPUT.dat};	
	\addplot+[blue,thick, dotted, no marks, select coords between index={5}{120}]	plot table[x=sample_avg,y=input_min]{AV10_W0_U_M4_D2_HB_S3_W2_10_R10_6_RP10_5_PB16384_HT512_ST1024_THROUGHPUT.dat};

	\nextgroupplot [title=($ht$:$2^{8}$),title style={font=\footnotesize,xshift=0, yshift=-3em}]
	\addplot+[red,thick, dotted, no marks, select coords between index={5}{120}]	plot table[x=sample_avg,y=input_max]{AV10_W0_U_M4_D2_HB_S3_W2_10_R10_6_RP10_5_PB16384_HT256_ST1024_THROUGHPUT.dat};
	\addplot+[black,very thick, no marks, select coords between index={5}{120}] plot table[x=sample_avg,y=input_avg]{AV10_W0_U_M4_D2_HB_S3_W2_10_R10_6_RP10_5_PB16384_HT256_ST1024_THROUGHPUT.dat};	
	\addplot+[blue,thick, dotted, no marks, select coords between index={5}{120}]	plot table[x=sample_avg,y=input_min]{AV10_W0_U_M4_D2_HB_S3_W2_10_R10_6_RP10_5_PB16384_HT256_ST1024_THROUGHPUT.dat};

	\nextgroupplot [title=($ht$:$2^{7}$),title style={font=\footnotesize,xshift=0, yshift=-3em}]
	\addplot+[red,thick, dotted, no marks, select coords between index={5}{120}]	plot table[x=sample_avg,y=input_max]{AV10_W0_U_M4_D2_HB_S3_W2_10_R10_6_RP10_5_PB16384_HT128_ST1024_THROUGHPUT.dat};
	\addplot+[black,very thick, no marks, select coords between index={5}{120}] plot table[x=sample_avg,y=input_avg]{AV10_W0_U_M4_D2_HB_S3_W2_10_R10_6_RP10_5_PB16384_HT128_ST1024_THROUGHPUT.dat};	
	\addplot+[blue,thick, dotted, no marks, select coords between index={5}{120}]	plot table[x=sample_avg,y=input_min]{AV10_W0_U_M4_D2_HB_S3_W2_10_R10_6_RP10_5_PB16384_HT128_ST1024_THROUGHPUT.dat};

	\nextgroupplot [title=($ht$:$2^{6}$),title style={font=\footnotesize,xshift=0, yshift=-3em}]
	\addplot+[red,thick, dotted, no marks, select coords between index={5}{120}]	plot table[x=sample_avg,y=input_max]{AV10_W0_U_M4_D2_HB_S3_W2_10_R10_6_RP10_5_PB16384_HT64_ST1024_THROUGHPUT.dat};
	\addplot+[black,very thick, no marks, select coords between index={5}{120}] plot table[x=sample_avg,y=input_avg]{AV10_W0_U_M4_D2_HB_S3_W2_10_R10_6_RP10_5_PB16384_HT64_ST1024_THROUGHPUT.dat};	
	\addplot+[blue,thick, dotted, no marks, select coords between index={5}{120}]	plot table[x=sample_avg,y=input_min]{AV10_W0_U_M4_D2_HB_S3_W2_10_R10_6_RP10_5_PB16384_HT64_ST1024_THROUGHPUT.dat};

	\nextgroupplot [title=($ht$:$2^{5}$),title style={font=\footnotesize,xshift=0, yshift=-3em}]
	\addplot+[red,thick, dotted, no marks, select coords between index={5}{120}]	plot table[x=sample_avg,y=input_max]{AV10_W0_U_M4_D2_HB_S3_W2_10_R10_6_RP10_5_PB16384_HT32_ST1024_THROUGHPUT.dat};
	\addplot+[black,very thick, no marks, select coords between index={5}{120}] plot table[x=sample_avg,y=input_avg]{AV10_W0_U_M4_D2_HB_S3_W2_10_R10_6_RP10_5_PB16384_HT32_ST1024_THROUGHPUT.dat};	
	\addplot+[blue,thick, dotted, no marks, select coords between index={5}{120}]	plot table[x=sample_avg,y=input_min]{AV10_W0_U_M4_D2_HB_S3_W2_10_R10_6_RP10_5_PB16384_HT32_ST1024_THROUGHPUT.dat};	
	
	\end{groupplot}
	
	\end{tikzpicture}	
	\vspace{-4mm}		
	\caption{Input throughput measurements ($w:2^{10}$).}
	\label{fig:throughput_time}
\end{figure}
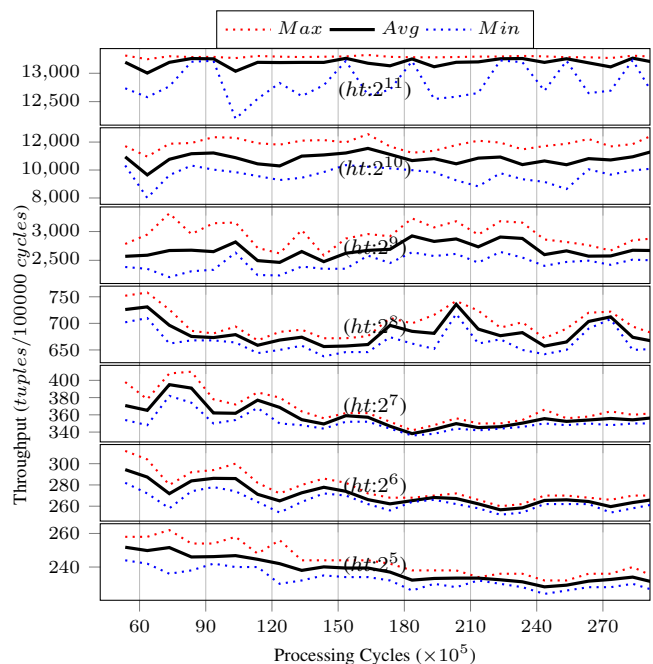

%% file: q3_power.tex
\begin{figure*}
	\centering
	\begin{subfigure}[t]{0.19\linewidth}
		\centering
		\includegraphics[width=1\linewidth]{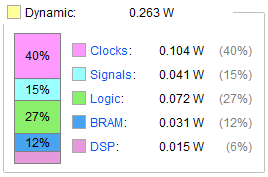}
				\vspace{-3mm}				
		\caption{10 nanoseconds.}
		\label{fig:q3_p10}	
	\end{subfigure}%
	\hfil
	\begin{subfigure}[t]{0.19\linewidth}
		\centering
		\includegraphics[width=1\linewidth]{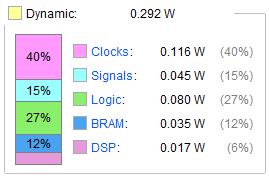}	
				\vspace{-3mm}			
		\caption{9 nanoseconds.}
		\label{fig:q3_p9}		
	\end{subfigure}
	\hfil
	\begin{subfigure}[t]{0.19\linewidth}
		\centering
		\includegraphics[width=1\linewidth]{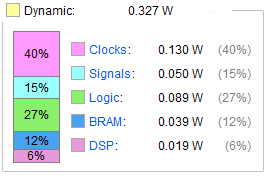}	
				\vspace{-3mm}			
		\caption{8 nanoseconds.}
		\label{fig:q3_p8}		
	\end{subfigure}
	\hfil
	\begin{subfigure}[t]{0.19\linewidth}
		\centering
		\includegraphics[width=1\linewidth]{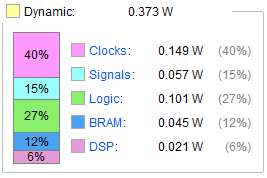}		
				\vspace{-3mm}		
		\caption{7 nanoseconds.}
		\label{fig:q3_p7}		
	\end{subfigure}
	\hfil
	\begin{subfigure}[t]{0.19\linewidth}
		\centering
		\includegraphics[width=1\linewidth]{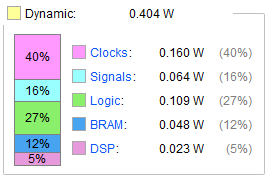}
		\vspace{-3mm}		
		\caption{6.5 nanoseconds.}
		\label{fig:q3_p65}		
	\end{subfigure}
	\caption{Clock frequency period effect on the power consumption of various system parts.}
	\label{fig:q3_power}
\end{figure*}

%% file: q3_util.tex
\begin{table*}
	\centering
	\caption{TPC-H third Query on \newfqp{} Resource Utilization (Virtex UltraScale XCVU190-2FLGC2104E FPGA)}
	\vspace{-3mm}
	\label{tab:tpchq3_table}
	\scriptsize
	\begin{tabular}{lccccccccccc}
		\toprule
		Unit          &    CLB     &     CLB     &    CLB    & CARRY8 & F7 Muxes & F8 Muxes &     LUT     &  LUT   &       LUT       & Block RAM & DSPs \\
		              &            &    LUTs     & Registers &        &          &          &    Logic    & Memory & Flip Flop Pairs &   Tile    &      \\
		Total         &   134280   &   107424    &  2148480  & 13420  &  537120  &  268560  &   1074240   & 231840 &     1074240     &   3780    & 1800 \\ \midrule
		FQP-Q3       &    9221    &    20695    &   40985   &  251   &   3193   &   1580   &    15947    &  4748  &      4183       &     6     & 130  \\
		-Brick1       &     98     &     499     &    318    &   0    &    0     &    0     &     155     &  344   &       16        &     0     &  0   \\
		-Brick2       &    8281    &    16678    &   37858   &  240   &   2944   &   1456   &    13582    &  3096  &      3890       &     6     & 128  \\
		--Input\_FIFO &     13     &     68      &     4     &   0    &    0     &    0     &     12      &   56   &        2        &     0     &  0   \\
		--LSwitch     &     2      &     11      &     0     &   0    &    0     &    0     &      3      &   8    &        0        &     0     &  0   \\
		--CMJoin      &    8221    &    16197    &   37831   &  240   &   2944   &   1456   &    13445    &  2752  &      3872       &     0     & 128  \\
		---PipeStage1 &    2054    &    3963     &   9501    &   60   &   736    &   364    &    3301     &  662   &       946       &    1.5    &  32  \\
		---PipeStage2 &    4081    &    7867     &   18429   &  120   &   1472   &   728    &    6655     &  1212  &      1852       &     3     &  64  \\
		---PipeStage3 &    2081    &    4037     &   9477    &   60   &   736    &   364    &    3375     &  662   &       951       &    1.5    &  32  \\
		---Auxiliary  &    147     &     330     &    303    &   0    &    0     &    0     &     114     &  216   &        9        &     0     &  0   \\
		--Filters     &     46     &     240     &    16     &   0    &    0     &    0     &     16      &  224   &        8        &     0     &  0   \\
		--Collector   &     34     &     162     &     7     &   0    &    0     &    0     &     106     &   56   &        4        &     0     &  0   \\
		-Brick3       &    262     &     695     &   1074    &   9    &   121    &    60    &     631     &   64   &       75        &     0     &  2   \\
		-Brick4       &    461     &    1180     &   1690    &   2    &   128    &    64    &     836     &  344   &       152       &     0     &  0   \\
		-GSwitch1-5   & $\sim$5*52 & $\sim$5*275 &    5*9    &   0    &    0     &    0     & $\sim$5*151 & 5*124  &    $\sim$5*5    &     0     &  0   \\
	\end{tabular}
    \vspace{-4mm}
\end{table*}

%% file: conclusions.tex
\section{Conclusions}\label{sec:conclusion}

Motivated by emerging IoT applications, interest in real-time stream
processing has increased in recent years.  In this work, we focused on
designing and developing a flexible hardware-based streaming
architecture with the capability of offline and online configuration
and adaptation to query workloads and their updates. This flexibility
is essential to many streaming applications, such as real-time data
analytics, information filtering, and complex event processing. The
proposed \newfqp\ framework allows us to choose a favorable
configuration for a set of queries. These queries can subsequently be
dynamically modified and adjusted online based on processing component
choices and their re-programmability properties. Furthermore, the
modular design of our processor, specifically the separation of the
unidirectional distribution network from the processing components,
simplifies the addition of other necessary computational components
(e.g., the query operators).